\documentclass[prb,aps,showpacs,groupedaddress,twocolumn,floatfix]{revtex4}
\usepackage{amsmath}
\usepackage{bm}
\usepackage{graphics}
\usepackage{dcolumn}
\begin{document}
\title{Spin dynamics in InAs-nanowire quantum-dots coupled to a transmission line}
\author{Mircea Trif, Vitaly N. Golovach, and Daniel Loss}
\affiliation{Department of Physics, University of Basel, 
Klingelbergstrasse 82, CH-4056 Basel, Switzerland}
\begin{abstract}
We study theoretically electron spins in nanowire quantum dots
placed inside a transmission line resonator. Because of the
spin-orbit interaction, the spins couple to the electric component
of the resonator electromagnetic field and enable coherent
manipulation, storage, and read-out of quantum information in an
all-electrical fashion. Coupling between distant quantum-dot spins,
in one and the same or different nanowires, can be efficiently
performed via the resonator mode either in real time or through
virtual processes. For the latter case we derive an effective
spin-entangling interaction and suggest means to turn it on and off.
We consider both transverse and longitudinal types of nanowire
quantum-dots and compare their manipulation timescales against the
spin relaxation times. For this, we evaluate the rates for spin
relaxation induced by the nanowire vibrations (phonons) and show
that, as a result of phonon confinement in the nanowire, this rate
is a strongly varying function of the spin operation frequency and
thus can be drastically reduced compared to lateral quantum dots in
GaAs. Our scheme is a step forward to the formation of hybrid
structures
where qubits of different nature can be integrated in a single
device.
\end{abstract}

\date{\today}
\draft{}
\preprint{}

\maketitle

%
\section{Introduction}%

\begin{figure}[t!]
\scalebox{0.173}{\includegraphics{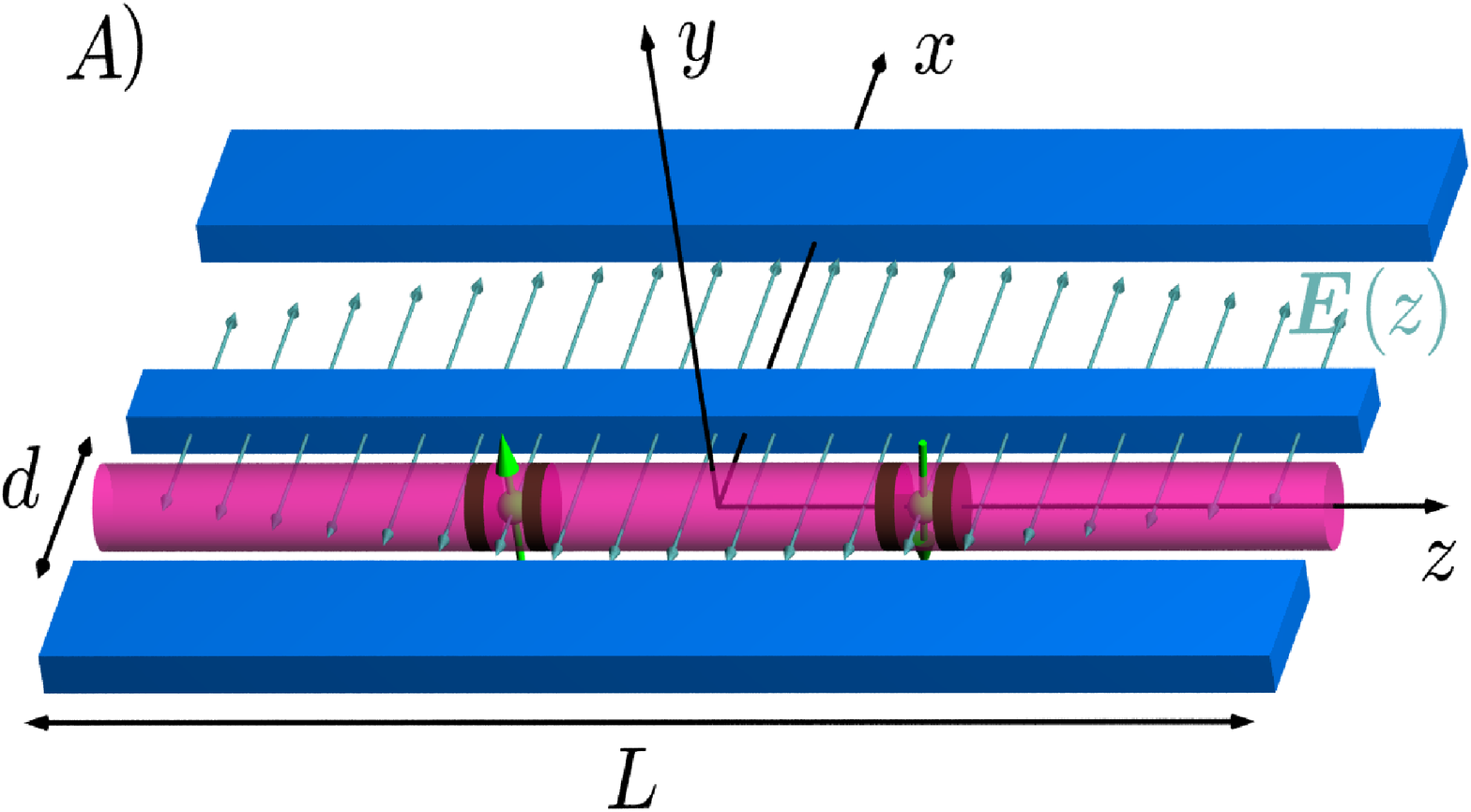}}\\

\scalebox{0.168}{\includegraphics{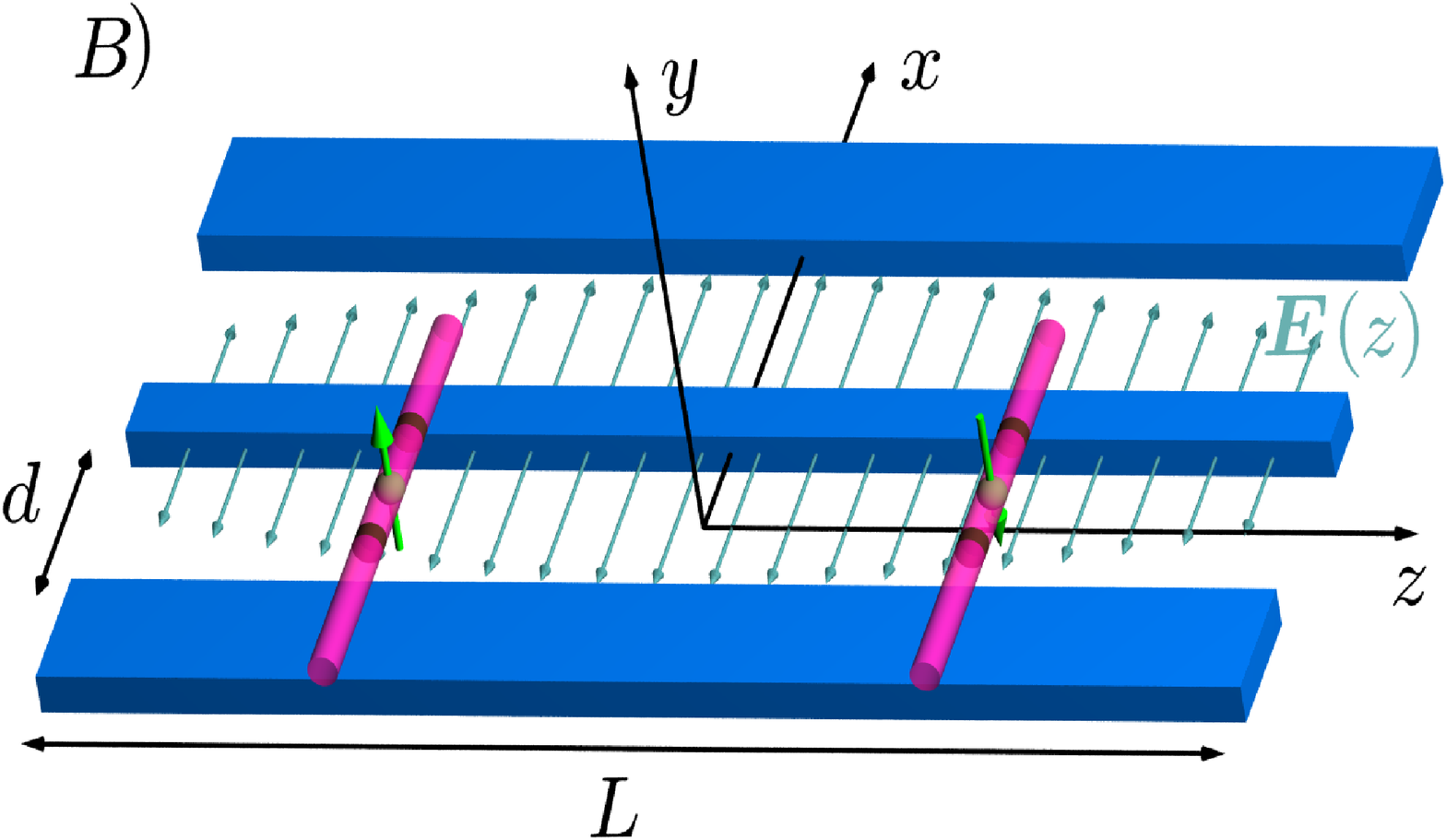}} \caption{
Schematics of the two configurations  considered in this work. A) 
Large-diameter InAs nanowire (pink) positioned inside and parallel to 
the transmission line (blue). 
The disk-shaped quantum dots (QD) are  located in the nanowire 
and are formed by two InP-boundaries (brown).
 Each QD contains
only one electron with spin 1/2 (green). B)
Two small-diameter InAs nanowires (pink) positioned
perpendicularly to the transmission line (blue).
The elongated  QDs  are oriented along the nanowire 
with one electron (green) in each dot. The QD confinement can be achieved
by  barrier materials (as shown in brown) or
by external gates (not shown).} \label{Fig1}
\end{figure}

Over the last decade, the spin of individual electrons
in semiconductor nanostructures has been intensively
studied in relation to  spin-based quantum
computing schemes.\cite{LDV,Cerletti,Hanson2007}
Attaining an almost full control over the spin of individual
electrons in QDs  opens the possibility to study single
spin dynamics in a solid state environment in the presence of
relaxation and decoherence.
Although lateral QDs have been most successfully used until
now to demonstrate spin coherence and usability for quantum
computing,\cite{Cerletti,Hanson2007} novel quantum systems have
emerged in recent years, providing a number of new ways to implement
the basic ideas of quantum computing.\cite{NielsenChuang} Among such
systems are the QDs formed inside semiconductor
nanowires.\cite{Fasth2007,Fuhrer}

Rapid progress in GaAs nanostructures started once few-electron QDs became
available (for a review, see e.g., Ref.\onlinecite{KAT}), which opened the door
to  control the number of electrons in a single QD
down to one  in  vertical\cite{Tarucha1996} and
lateral\cite{Ciorga2000} dots, as well as in double QDs.\cite{Elzerman2003,Hayashi2003, Petta2004}
Further important experimental progress came with the advent of charge sensors which, quite
remarkably, enabled 
 the
measurement of the relaxation time of one single spin.\cite{Elzerman2004}
The longest spin relaxation times in single GaAs QDs extend up to several seconds\cite{Amasha2006}
and were measured in lateral dots at relatively small magnetic fields ($B\sim 1\,{\rm T}$).

The spin decoherence time in GaAs was measured also in double QDs
by studying the hyperfine-induced
mixing of 
singlet and triplet states.\cite{JohnsonPettaT2Nature,Petta2005} In the same set-up, a universal 
entanglement operation was implemented,\cite{Petta2005} enabling a square-root-of-swap 
operation\cite{LDV} between two spin-1/2 qubits  on a time scale of 180 ps.
Resonant and coherent manipulation of a single spin-1/2  has recently been implemented  in a GaAs 
double QD, making use of electron spin resonance (ESR)\cite{Engel,Koppens2006} as well as 
electric dipole induced spin resonance (EDSR)\cite{Golovach2006,Nowack}
techniques. Resonant but incoherent (hyperfine-mediated) spin manipulation  in double dots was also 
recently demonstrated.
\cite{Laird07}

The use of different semiconductors, other than GaAs, has since long
been pursuit with the goal to create nanostructures with novel
properties. Particular examples are InAs and InP nanowires, where
both gate defined and 'barrier' defined QDs could be
fabricated.\cite{Bjork2004,Fasth2005,Pfundt2006,Shorubalko2007}
The advantage of these materials is that both optical and transport
measurements can be carried out on the same type of structure. The
number of electrons can equally well be controlled down to one
electron per dot,\cite{Fasth2005} which shows that 
QDs created in nanowires can serve as alternative candidates for
spin-qubits.

One particular difference between GaAs and InAs semiconductors is the
strength of the spin-orbit interaction (SOI), which is much larger
for the latter material. This fact, however, is  a double-edge sword; on one hand it
opens up the possibility to efficiently manipulate the electron spin
with {\em electric} fields only,\cite{Rashba2003, Kato2004, Duckheim2006,
Golovach2006, Trif2007} while on the other hand it implies stronger
coupling of the spin to charge environments, like phonons,
particle-hole excitations, gate voltage fluctuation, etc. However,
due to the quasi-1D structure of the nanowires, the spin relaxation
times due to phonons and SOI turn out to be longer than one might expect from QDs created
in InAs bulk material. Indeed, the time scales  obtained in this
work  are  on the order of microseconds to milliseconds
for sufficiently large Zeeman splittings. At the same time, the relaxation
rate exhibits peaks as a function of a static applied magnetic field due
to the quantization of the phonon spectrum. The long relaxation time
and the presence of strong SOI permits then an efficient  control of coherent spin states
by making use of EDSR. \cite{Rashba2003, Kato2004, Duckheim2006,
Golovach2006,Nowack}

One of the main ingredients in the spin-qubit scheme\cite{LDV} is the electrical
control of
two-qubit gates
to generate entanglement.  While the original proposal involved only local
interactions between neighboring spins, 
 it is desirable  to couple  spins directly
over large distances, since this produces a better threshold for fault tolerant quantum 
computation.\cite{Terhal} A solution to this problem was first proposed
in Ref. \onlinecite{Imamoglu} and involves  optical cavities
whose photon modes  mediate  interaction between distant spins. The
coupling of the spin  to optical cavities in semiconductors was also the subject of
some recent experiments.\cite{Berezovsky2006,Atature2007}

Very recently, 1D electromagnetic cavities (or transmission lines) were shown
to be very suitable for reaching the strong coupling regime between
superconducting qubits and photons.\cite{Wallraff2004,Blais2004,Gywat}
Theoretical extension to QDs were proposed subsequently, including charge and
spin qubits.\cite{Lukin2004,Burkard2006} The direct coupling of
the spin to the cavity modes via the magnetic dipole transitions is
usually weak and one has to use electric dipole transitions together
with correlations between spin and charge degrees of freedom in
order to obtain a strong effective coupling.  This can be achieved
in several ways, e.g. by making use of  the Pauli exclusion
principle and Coulomb repulsion, \cite{Burkard2006} or  of Raman
transitions.\cite{Lukin2004}

Here we propose  another mechanism to achieve long-distance  coupling
between spins inside a cavity, 
namely via SOI which leads to an effective coupling of spin to the electric
field component of the cavity photon, and thus eventually to a  coupling
between distant spins mediated by this photon. In order to reach a sizable
coupling strength, it is desirable to use nanostructures with large SOI  such as
 InAs QDs. Two such proposed configurations, which define the two model systems
 to be studied in this paper, are sketched in Figs. \ref{Fig1} A and B. They 
 consist of nanowire  QDs  embedded
 in  a transmission line. In particular, in Fig. \ref{Fig1}A a nanowire
 positioned {\em parallel}
to the transmission line axis is shown. 
In this case, the QDs
are realized by confining the electrons in the longitudinal
direction (i.e. along the nanowire axis) much stronger than in the transverse one. This corresponds
to a nanowire with a large diameter, on the order of $80-100\,{\rm
nm}$. Such longitudinal  confinement  can be achieved
by applying metallic gates or  by using other materials as barriers (InP for example,
which is depicted in Fig. \ref{Fig1} in brown) which have a larger band
gap than the host material such as e.g. InAs.\cite{Fasth2005,Fuhrer} In Fig
\ref{Fig1}B a small-diameter ($D<40\,{\rm nm}$) InAs nanowire is shown, being
positioned perpendicularly to the transmission line and containg  QDs that are elongated
along the nanowire. That means that in this case we assume that
the electronic confinement along  the nanowire is  much weaker than
in the transverse direction. Then, to a very good approximation, the
electrons can be considered as behaving one-dimensionally,
which will allow us to treat the SOI exactly, while this is not possible for 
the configuration Fig. \ref{Fig1}A.
However, in order to prevent a current flow, the nanowire and the
transmission line need to be separated by some insulating coating material,
obtained, for example, by atomic layer deposition.

The goal of our work is now to analyze these configurations in detail and,
in the first part of the paper,
to derive  an effective spin-spin coupling Hamiltonian. In the second 
part, we study the spin decay in this system, induced by
phonons and SOI, and calculate explicitly the spin relaxation and decoherence
 times due to this mechanism. We will show that these times are
 much longer than the switching times needed to manipulate and
 couple the spins coherently. Thus, our findings provide theoretical evidence that
 nanowire QDs embedded into transmission lines are promising candidates
 for spin-qubits with tunable long-range coupling. This scheme also
 opens the door to hybrid configurations where  qubits of different nature (e.g.
 superconducting and spin qubits) can be coupled via the transmission line.

The paper is organized as follows. In Sec. II we introduce the model
for our system, namely single-electron QDs and cavity and specify
the model Hamiltonian. In Sec. III we derive first the effective
spin-photon Hamiltonian for a single spin in the cavity for a
general SOI. Here  we derive also the general effective
spin-spin coupling induced by the SOI and  the cavity photon modes. In Sec.
IV we investigate the case of a QD strongly confined in the
longitudinal direction. Then, in Sec. V we analyze the opposite
case, when the electron is strongly confined in the transverse
direction of the nanowire. In Sec. VI we provide some numerical
estimates for the strengths of the spin-photon and spin-spin
couplings for both cases. Then, in  Sec. VII we give a brief
description of the manipulation of the spins by electric fields. In
Sec. VIII we study the spin decay and provide a detailed description
of the relaxation of the spin via SOI and acoustic phonons. Some
technical details of the phonon analysis are deferred to App. A.
Finally, conclusions are given in Sec. IX.

\section{Model Hamiltonian}%

The Hamiltonian of the system composed of the single-electron QD and the cavity reads
\begin{equation}
H=\frac{p^2}{2m^{*2}}+V(\bm{r})+\frac{1}{2}
g\mu_B\bm{B}\cdot\bm{\sigma}+H_{SO}+H_{e-\gamma}
+H_{\gamma},
\label{hamiltonian}
\end{equation}
where the first two terms represent the  bare orbital part of the
Hamiltonian, $m^*$ is the effective
mass of the electron, $g$ is the
$g$-factor of the electron in the material, and $V(\bm{r})$ is the
confinement potential, both in the longitudinal and transverse
directions. We can obtain an effective Hamiltonian $H_{\it eff}$ by
averaging over the ground-state $|0\rangle$ in the longitudinal or
in the transverse directions  depending on which case in Fig.
\ref{Fig1} is considered. Then, for the system in Fig.
\ref{Fig1}A(B) we obtain an effective 2D (1D) Hamiltonian.

The third term stands for the Zeeman interaction, while the fourth
term in Eq. (\ref{hamiltonian}) represents the SOI. For wurtzite
InAs nanowires grown along the $c$-axis, with the longitudinal
confinement much stronger than the transverse one (see Fig.
\ref{Fig1}A) the SOI takes the form of a Rashba
type,\cite{Fasth2007} $H_{SO}\equiv
H_{SO}^{t}=\alpha(\bm{p}\times\bm{c})\cdot\bm{\sigma}$, which, when
written in components, becomes
\begin{equation}
H_{SO}^t=\alpha(p_x\sigma_y-p_y\sigma_x)\,. \label{spinorbittrans}
\end{equation}
We mention that our present study is quite general and can be
easily adapted to other types of SOIs (such as Dresselhaus type).
In the opposite case, when the transverse confinement is much
stronger than the longitudinal one (see Fig. \ref{Fig1}B), the SOI
Hamiltonian $H_{SO}$ takes the form $H_{SO}\equiv
H_{SO}^l=(\bm{k}\cdot\bm{c})(\bm{\eta}\cdot\bm{\sigma})$ which, when
written in components, becomes
\begin{equation}
H_{SO}^l=\eta p_x\sigma_{\bm{\eta}},\label{spinorbitlong}
\end{equation}
with $\bm{\eta}=(\eta_x,\eta_y,\eta_z)$ being a vector of coupling
constants and $\sigma_{\bm{\eta}}$ being the spin component along
$\bm{\eta}$.\cite{Fasth2007}

The sixth term represents the interaction between the photons in the
cavity, labeled  $\gamma$, and the electron in the QD. This term
is given  by
\begin{equation}
H_{e-\gamma}=e\bm{E}(z)\cdot\bm{r}.\label{ephoton}
\end{equation}
The  electric field $\bm{E}(z)$ acting on the electron is
$\bm{E}(z)=\bm{e}_y\,V(z)/d$, with $\bm{e}_y$ being the unit vector along
$y$, $V(z)$ represents the fluctuating potential within the
transmission line and $d$ is the distance between the transmission line and the
center conductor. The voltage fluctuation $V(z)$  has the following
form \cite{Blais2004}:
\begin{equation}
V(z)=\sum_{p=1}^{\infty}V_p\sin{\left(\frac{p\pi z}{L}\right)}[a_p+a_p^{\dagger}], \label{1}
\end{equation}
where $V_p=\sqrt{\hbar\omega_p/Lc}$, $a^{\dagger}_p$($a_p$) are the
creation (annihilation) operators for the excitations(photons), $c$
the capacitance per unit length, $L$ the legth of the resonator,
and $\omega_p$ the eigenmodes of the resonator. The last term in the
Hamiltonian represents  the free photons
$H_{\gamma}=\sum_{p}\hbar\omega_pa^{\dagger}_pa_p$.

From Eq. (\ref{hamiltonian}) we see that there exists an  infinite
number of frequencies in the transmission line,  implying a coupling
of the electron charge to an infinite number of modes. However, from
all these modes, the relevant ones are  those close to resonance
with the Zeeman splitting of the spin. In the following  we
disregard all other modes from the problem and we assume also that
the QD is in the center of the transmission line, so that the
interaction between the electron charge and the photons becomes
maximal.
Having now
defined all the ingredients, we can proceed to study the dynamics of
the system.

\section{General Spin-photon dynamics}%

\subsection{Spin-photon interaction}%

In the following we derive  an effective spin-photon Hamiltonian,
assuming for both cases in Fig. \ref{Fig1} a SOI of arbitrary strength
(to be restricted later on).
 In the case of time-reversal symmetry, the ground
state of the dot ($H_d\equiv H_0+H_{SO}+H_Z$) is two-fold degenerate
(Kramers doublet), while this degeneracy is lifted in  the presence
of a magnetic field. If the magnetic field is such  that the doublet
splitting  and also the electron-photon coupling strength are
smaller than the level spacing of the QD, we can restrict our considerations
to the dynamics of the lowest doublet only. We label this doublet by
$\{|\Uparrow\rangle,|\Downarrow\rangle\}$, which is now  different
from the 'true' electron spin.

We can connect formally the states in the presence of the SOI to the
ones in the absence of the SOI  with the help of a unitary
transformation or Schrieffer-Wolff (SW) transformation
\begin{equation}
|n_\tau\rangle=e^{-S}|n\rangle|\sigma\rangle,\label{transformation}
\end{equation}
where the states $|n\rangle$ are the eigenstates  of the
Hamiltonian $H_0$ ($H_0|n\rangle=E^0_n|n\rangle)$, $|n_\tau\rangle$
are the Kramers doublets with SOI,
$|\sigma\rangle=|\uparrow,\downarrow\rangle$ are the bare spin
states, and $S=-S^{\dagger}$. Also, the relation
$H_d|n_\tau\rangle=E_d^{n_\tau}|n_\tau\rangle$ holds from our
definition of the transformed state. For notational convenience we
denote the lowest Kramers doublet as  $|0_\tau\rangle$. This is
written simply as $|0_\tau\rangle\equiv|\tau\rangle$, with the
identification
$|\tau\rangle=\{|\Uparrow\rangle,|\Downarrow\rangle\}$. The above
transformation can be performed on the level of the Hamiltonian,
implying diagonalization of  the Hamiltonian $H_d$ in the basis of
the 'bare' Hamiltonian $H_0$
\begin{equation}
\bar{H}\equiv e^{-S}He^S.
\end{equation}
The advantage of
transforming the Hamiltonian $H_d$ so  that it becomes diagonal in
the basis of the bare Hamiltonian $H_0$ is now obvious. Within this
transformation one can  in principle proceed to calculate the effect
of SOI  to arbitrary order in perturbation theory, together with the
SOI induced spin-photon coupling. We can now derive an effective
spin-photon Hamiltonian within the lowest doublet $|\tau\rangle$ by
averaging $\bar{H}$ over the orbital ground state $|0\rangle$. This
leaves us with the following effective spin-photon Hamiltonian
$H_{s-\gamma}\equiv\langle0|\bar{H}|0\rangle$ given
by
\begin{equation} H_{s-\gamma}=\frac{1}{2}g\mu_B
B_{\it{eff}}\sigma_z+\bm{\mathcal{M}}_{\it{\gamma}}
\cdot\bm{\sigma}(a^{\dagger}+a)+\hbar\omega
a^{\dagger}a,\label{effective}
\end{equation}
where
\begin{equation}
\frac{1}{2}g\mu_BB_{{\it eff}}\sigma_z=\langle0|e^{-S}
H_de^{S}|0\rangle\label{Zeemanrenorm}
\end{equation}
 stands for the renormalized magnetic field  and
\begin{equation}
\bm{\mathcal{M}}_{\gamma}\cdot\bm{\sigma}=\frac{eV_0}{d}
\langle0|e^{-S}ye^{S}|0\rangle.
\label{spinphnon1}
\end{equation}
stands for the spin-photon coupling. We mention that in order to
have a finite coupling of  the spin $\bm{\sigma}$ to the photons,
the vector $\bm{\mathcal{M}}_{\gamma}$ must contain some
time-reversal breaking parameter, such as the external magnetic
field $B$. In  the absence of the magnetic field there is no
coupling  between the lowest doublet and the photons
($\bm{\mathcal{M}}_{\gamma}=0$) to all orders in SOI.

We now define the spin-photon coupling strength
$\nu=\sqrt{(\mathcal{M}_{\gamma}^{x})^2+(\mathcal{M}_{\gamma}^{y})^2}$
and the detuning of the qubit from the cavity by 
$\Delta\equiv E_{Z}^{{\it eff}}-\hbar\omega$, where $E_Z^{\it eff}\equiv
g\mu_BB_{\it eff}$. Close to the resonance between the qubit and the
cavity mode ($\Delta\ll E_{Z}^{{\it eff}},\hbar\omega$) one can
simplify Eq. (\ref{effective}) by  using the so called  rotating
wave approximation (RWA).\cite{RWA} This implies to switch first to
the interaction picture, so that the operators $a(a^{\dagger})$ and
$\sigma_{\mp}$, where $\sigma_{\mp}=\sigma_x\mp i\sigma_y$ become
time-dependent
\begin{eqnarray}
\sigma_{\mp}(t)&=&\sigma_{\mp}(0)e^{\mp i\omega_Z^{{\it eff}}t}\\
a(t)&=&a(0)e^{-i\omega t}\\
\sigma_z(t)&=&\sigma_z(0).\label{intpicture}
\end{eqnarray}
where  $\omega_Z^{\it eff}=E_{Z}^{\it eff}/\hbar$. Then, we neglect
the terms in the time-dependent resulting Hamiltonian which
oscillate fast on the time scale $\hbar/\Delta$. This means
neglecting counter-rotating terms such as $a^{\dagger}\sigma_+\sim
e^{i(\omega_Z^{{\it eff}}+\omega)t}$, $a\sigma_-\sim
e^{-i(\omega_Z^{{\it eff}}+\omega)t}$, $a^{\dagger}\sigma_z\sim
e^{i\omega t}$, and $a\sigma_z\sim e^{-i\omega t}$, which average to
zero for large times. Within this approximation  the Hamiltonian in
Eq. ($\ref{effective}$) within the interaction picture becomes
static and of the form
\begin{equation}
H_{s-\gamma}^{{\it eff}}=\frac{1}{2}g\mu_BB_{{\it eff}}
\sigma_z+\nu(a^{\dagger}\sigma_-+\sigma_+a)
+\hbar\omega a^{\dagger}a.\label{JC}
\end{equation}
As expected, the above expression agrees with the Jaynes-Cummings
Hamiltonian.\cite{JaynesCummings}

\subsection{Effective spin-spin interaction}%

We now investigate the case of two QDs in the cavity in the limit of finite 
detunings $\Delta_{1,2}$. 
The Hamiltonian $H^{(2)}_{s-\gamma}$ corresponding to  the two spins 
in the cavity  can be found by just 
extending Eq. (\ref{JC}) to two spins
\begin{equation}
H^{(2)}_{s-\gamma}=\sum_{i=1,2}\left(
\frac{1}{2}g_i\mu_BB^i_{{\it eff}}\sigma_z^i+
\nu_i(a^{\dagger}\sigma_-^i+\sigma_+^ia)\right)+
\hbar\omega a^{\dagger}a.\label{JCtwospins}
\end{equation}
For $\nu_i/\Delta_i<1\,(i=1,2)$,  the spin-photon interaction can be treated
within the second order perturbation theory in $\nu_i$. We use again the SW 
transformation, similar to the previous section. Here, this implies 
finding an  operator $T$ so that
\begin{equation}
\widetilde{H}^{(2)}_{s-\gamma}=e^TH^{(2)}_{s-\gamma}e^{-T}\label{newSW}
\end{equation}
is diagonal in the basis of the spin-photon Hamiltonian without  spin-photon interaction 
(the Hamiltonian $H^{(2)}_{s-\gamma}$ with $\nu_{1,2}\equiv0$). 
Within  first order in spin-photon couplings 
$\nu_{1,2}$, the transformation operator $T$ reads
\begin{equation}
T=\sum_{i=1,2}\frac{\nu_i}{\Delta_i}(\sigma^i_+a-
a^{\dagger}\sigma^i_-),\label{transformation2}
\end{equation}
under the assumption  that the condition $\nu_i/\Delta_i<1$, $i=1,2$, 
is satisfied for both dots. The transformed 
Hamiltonian $\widetilde{H}^{(2)}_{s-\gamma}$ becomes
\begin{eqnarray}
\widetilde{H}^{(2)}_{s-\gamma}=\left(\hbar\omega+\frac{2\nu_1^2}{\Delta_1}
\sigma_z^1+\frac{2\nu_2^2}{\Delta_2}\sigma_
z^2\right)a^{\dagger}a\nonumber\\+\left(E_{1Z}^{{\it 
eff}}+\frac{\nu_1^2}{\Delta_1}\right)\sigma_z^1+\left(E_{2Z}^{{\it 
eff}}+\frac{\nu_2^2}{\Delta_2}\right)\sigma_z^2\nonumber\\
+\nu_1\nu_2\left(\frac{1}{\Delta_1}+\frac{1}{\Delta_2}\right)(\sigma_+^1\sigma_-^2
+\sigma_+^2\sigma_-^1),\label{SW}
\end{eqnarray}
where $E_{iZ}^{{\it eff}}=g\mu_BB^i_{{\it eff}}$. 
We can obtain a pure spin 
Hamiltonian by neglecting the 
fluctuations of the photon number $a^{\dagger}a\rightarrow \langle 
a^{\dagger} a\rangle\equiv \bar{n}$, with 
$\bar{n}$ the average number of photons in the lowest cavity mode. The resulting Hamiltonian 
$H_s\equiv\widetilde{H}^{(2)}_{s-\gamma}|_{a^{\dagger}a\rightarrow \bar{n}}$ reads
\begin{equation}
H_s=\widetilde{E}_Z^{1}\sigma_z^1+\widetilde{E}_Z^{2}\sigma_z^2+
J(\sigma_+^1 \sigma_-^2+\sigma_+^2 \sigma_-^1),\label{spin}
\end{equation}
where
\begin{equation}
\widetilde{E}_Z^i= E_{iZ}^{{\it eff}}+2\left(\bar{n}+\frac{1}{2}\right)
\frac{\nu_i^2}{\Delta_i},\;\;\;\;i=1,2,\label{renormZeeman}
\end{equation}
\begin{equation}
J=
\nu_1\nu_2\left( \frac{1}{\Delta_1}
+\frac{1}{\Delta_2}\right)
\label{spinspincoupling}.
\end{equation}
In Eq. (\ref{renormZeeman}) we see that the effective Zeeman 
splitting $\widetilde{E}_{Z}^{i}$ is quite different 
from the bare one, $E_{iZ}$. Besides the SOI renormalization of the Zeeman splitting, there is 
also a 
contribution from the spin-photon coupling, which consists of the Lamb shift (the term independent 
of the average 
photon number $\bar{n}$) and the ac Stark shift (the term 
proportional to the average photon number $\bar{n}$).

The expression  Eq. (\ref{spin}) is one of our main results: 
in the presence of SOI and cavity modes one can 
achieve an effective spin-spin coupling with the exchange coupling $J$  
between two  spins that are spatially 
well-separated. Indeed, this interaction  can act over the entire length 
of the cavity, which can be as large as 
a few millimeters.  Also, the spin-spin  interaction 
is of XY-type (transverse spin-spin coupling), which 
together with single spin rotations have been shown to be universal for quantum 
computing.\cite{Imamoglu,Blais2007} We mention that  in order to 
obtain a maximal effect, one should be able to 
tune the two qubits into resonance, so that $\widetilde{E}_Z^1=\widetilde{E}_Z^2$.\cite{Imamoglu}

\section{Strong longitudinal confinement}%


So far we have taken the SOI into account exactly, regardless of the  system under consideration, 
but  under the assumption that the lowest Kramers doublet is well separated from the higher states 
compared to Zeeman energy and electron-photon coupling. We analyze here the spin-photon coupling 
for the case shown in Fig. \ref{Fig1}A.  As stated in Section II, in this case we can derive an 
effective transverse Hamiltonian $H_{\it eff}\equiv H_{t}=\langle0_l|H|0_l\rangle$, where 
$|0_l\rangle$ stands for the ground-state wave-function in the longitudinal direction $z$.  The 
effective  Hamiltonian $H_t$ reads
\begin{equation}
H_t=\frac{p_x^2+p_y^2}{2m^*}+V(x,y)+H_{Z}+H_{SO}^t+H_{e-\gamma}+H_{\gamma},\label{transverseham}
\end{equation}
with $V(x,y)=\langle0_l|V(\bm{r})|0_l\rangle$, while all the other terms stay the same since they 
do not act in the $z$-direction. In the above expression we disregarded the term 
$\langle0_l|(p_z^2/2m^*)|0_l\rangle$, as it gives  a constant shift of the levels.

We can start in principle to derive the spin-photon interaction from the 
effective Hamiltonian $H_t$ by making use of the transformation (\ref{transformation}). 
However, this cannot be done exactly and we have to proceed in perturbation theory. 
In order to give some numerical estimates for the strength  
of the coupling $\nu$, we assume  the limit of weak  SOI, quantified by the condition 
$R/\lambda_{SO}<1$, with $R$ being the dot (wire) radius 
and $\lambda_{SO}=\hbar/m^*\alpha$ the spin-orbit length.\cite{Vitaly,Trif2007} 
Then, we can  treat the SOI within  perturbation theory. We assume in the following 
hard-wall boundary conditions for the electrons  confined in the QDs, namely circular 
hard-wall boundaries in the transverse direction. In the longitudinal direction the 
electron is also confined by a hard-wall type of potential, but much stronger than 
in the transverse direction, as stated before. We compute the operator $S$ from 
Eq. (\ref{transformation}) within the first order  in 
SOI, $S\approx (L_0+L_Z)^{-1}H_{SO}$, which gives explicitly
\begin{equation}
S\approx i\bm{\xi}\cdot\bm{\sigma}-E_ZL_0^{-1}
(\bm{l}\times\bm{\xi})\cdot\bm{\sigma},\label{smatrix}
\end{equation}
in the limit of $E_Z<<\Delta E_0$ with $\Delta E_0=E_1-E_0$ being the energy 
difference between the first excited state $|1\rangle$ and  the ground state 
$|0\rangle$. In the above formulas the Liouvilleans $L_{0,Z}$ are defined 
as  $L_{0,Z}A=[H_{0,Z},A]$  $\forall A$ and $\bm{\xi}=\lambda_{SO}^{-1}(-y,x,0)$, 
$\bm{l}=\bm{B}/B$. We can obtain an effective Hamiltonian up to second-order in 
SOI and first order in Zeeman splitting for the lowest Kramers doublet by averaging 
over the orbital ground state $|0_t\rangle$,
\begin{eqnarray}
H_{s-\gamma}&=&\frac{1}{2}g\mu_B\bm{B}\cdot\bm{\sigma}
+\langle0|[S,H_{SO}]|0\rangle+\langle0|[S,H_
{e-\gamma}]|0\rangle\nonumber\\
&+&\frac{1}{2}\langle0|[S,[S,H_{e-\gamma}]]|0\rangle+H_{\gamma}.
\label{perturbativeham}
\end{eqnarray}
The orbital wave-functions have the form (for circular hard-wall boundary conditions)
\begin{equation}
\psi_{mp}(r)=\frac{1}{\sqrt{\pi}R}\frac{e^{im\phi}}{J_{|m|+1}(k_{mp}R)}J_{|m|}(k_{mp}r),\label{wavefunction}
\end{equation}
where $J_{|m|}(k_{mp}r)$ are the Bessel functions of the first kind, $r$ is the electron radial 
coordinate in the transverse direction, and $k_{mp}$ are the solutions of the equation
$J_{|m|}(k_{mp}R)=0$. The appropriate energies are given  by $E_{mp}=\hbar^2 k_{mp}^2/2m^*$. Also, 
we assume that the magnetic field $\bm{B}$ and  the fluctuating electric field $\bm{E}$ are both 
along the $y$ direction, such that $H_{e-\gamma}=eE\,y$ and 
$S=i\bm{\xi}\cdot\bm{\sigma}-(E_Z/\lambda_{SO})L_0^{-1}y\,\sigma_z$.  After performing  the 
integrations,  we are left with the following effective Hamiltonian
\begin{equation}
H_{s-\gamma}=\frac{1}{2}E_Z^{{\it eff}}\sigma_z+\mathcal{M}_{\gamma}^y(a^{\dagger}+a)\sigma_y+H_{\gamma},\label{spinphotonpert}
\end{equation}
with
\begin{equation}
E_Z^{{\it eff}}\simeq
E_Z\left(1-0.25\left(\frac{R}{\lambda_{SO}}\right)^2\right),\label{Zeemaneffective}
\end{equation}
\begin{equation}
\mathcal{M}_{\gamma}^y\simeq0.25eE\,R\frac{E_Z}{\Delta E_0}\frac{R}{\lambda_{SO}}.\label{speffective}
\end{equation}
We see that there is no second order  contribution in SOI to the spin-photon interaction, this 
contribution vanishes identically for cylindrical wires in the ground state. We mention that within 
the RWA the Jaynes-Cummings coupling $\nu$ becomes $\nu=\mathcal{M}_{\gamma}^y$.

In the case of two spins present in the cavity, one obtains the same expression as in Eq. 
(\ref{spin}), where  $\nu_{1,2}$ is given by Eq. (\ref{speffective}). Since our coupling is 
proportional to the bare Zeeman splitting $E_Z$, we need large magnetic fields in order to obtain a 
strong coupling. Then, we can in principle neglect the Lamb and the ac Stark shifts in the 
expressions for $\widetilde{E}^i_Z$, since they give negligible renormalizations, so that 
$\widetilde{E}^i_Z\approx E^{{\it eff}}_{iZ}$. However, as can be seen from Eq. 
(\ref{Zeemaneffective}), the Zeeman splitting can be strongly reduced for large  SOI. This feature 
will turn out to be very important in order to have a long-lived qubit (see below).

\section{Strong transverse confinement}%

In this section we analyze the case shown in Fig. \ref{Fig1}B, i.e. when the transverse confinement 
is much stronger than the longitudinal one. As in the previous case, we can derive an effective 
longitudinal Hamiltonian by averaging the full Hamiltonian $H$ over the transverse orbital 
ground-state $|0_t\rangle$. The effective Hamiltonian $H_{\it eff}\equiv 
H_l=\langle0_t|H|0_t\rangle$ reads
\begin{equation}
H_l=\frac{p_x^2}{2m^*}+V(x)+H_Z+H_{SO}^l+H_{e-\gamma}+H_{\gamma},\label{longhamiltonian}
\end{equation}
with $V(x)=\langle0_t|V(\bm{r})|0_t\rangle$, while  all other terms remain the same, since they 
have no  action along the  $x$-direction. Again, like in the previous case, we disregard the term 
$\langle0_t|(p_y^2+p_z^2)/2m^*|0_t\rangle$, since it gives a constant shift of the levels.

We now derive the spin-photon interaction from the effective Hamiltonian (\ref{longhamiltonian}). 
As can be seen from Eq. (\ref{spinorbitlong}), the SOI contains only one spin-component, 
$\sigma_{\bm{\eta}}$ along the $\bm{\eta}$-direction. In this case and  in the absence of an 
external magnetic field  the SW transformation (\ref{transformation}) can be performed exactly, 
since the SOI appears as an Abelian gauge-potential.\cite{Rashba2001,Flindt2006} In the presence of 
an external magnetic field, however,  this cannot be done exactly anymore. We now apply the 
transformation (\ref{transformation}) to the Hamiltonian $H_l$ so that we obtain 
$\bar{H}_{l}=e^{-S}He^S$, with the operator $S$ corresponding to the zero-field case. This operator 
$S$ reads
\begin{equation}
S=-i\frac{x}{\lambda_{SO}}\sigma_{\bm{\eta}},\label{exacttransform}
\end{equation}
with $\lambda_{SO}=\hbar/m^*\eta$. The effect of this transformation can be evaluated exactly and 
we obtain
\begin{eqnarray}
\bar{H}_l = \frac{p_x^2}{2m^*}+V(x)+H_Z(x)+eEx+\hbar\omega a^{\dagger}a,\label{spinhamlong}
\end{eqnarray}
with
\begin{eqnarray}
H_{Z}(x)=\frac{1}{2}g\mu_B\bigg(\cos\left(\frac{2x}{\lambda_{SO}}\right)\bm{B}_{\bm{\eta}\perp}\cdot\bm{\sigma}\nonumber\\+B_{\bm{\eta}}\sigma_{\bm{\eta}}-\sin\left(\frac{2x}{\lambda_{SO}}\right)(\bm{e}_{\bm{\eta}}\times\bm{B})\cdot\bm{\sigma}\bigg),\label{Zeemanx}
\end{eqnarray}
where $\bm{B}_{\bm{\eta}\perp}$ is the component of the magnetic field
$\bm{B}$ perpendicular to the  vector $\bm{\eta}$, $B_{\bm{\eta}}$ is
the magnetic field component along $\bm{\eta}$, and
$\bm{e}_{\bm{\eta}}=\bm{\eta}/\eta$. We now assume, as before, that
the Zeeman splitting $E_Z=g\mu_BB$ is much smaller than the orbital
level spacing $\Delta E_0$ given by the first two term in the above
Hamiltonian. Also, we assume harmonic confinement potential along the
$x$-direction $V(x)=m^*\omega_0^2x^2/2$ which gives a dot size
$l=\sqrt{\hbar/m^*\omega_0}$. This is usually the case for
gate-defined QDs. Then, the above condition translates in having
$E_Z\ll\hbar\omega_0$. We are now in position to derive an effective
spin-photon Hamiltonian by treating $H_Z(x)$ within perturbation
theory. We perform a  new SW transformation and  transform the
above Hamiltonian into a diagonal one in the basis of $H_0$ to obtain 
$H_{s-\gamma}=\langle0|e^{-S'}\bar{H}e^{S'}|0\rangle$. We averaged also over the orbital ground 
state $|0\rangle$ to obtain a pure spin-photon Hamiltonian. Within lowest order in 
$E_Z/\hbar\omega_0$ the transformation is given by $S'=(1-\mathcal{P})L_0^{-1}H_Z(x)$. After 
inserting the operator $S'$ in the expression for $H_{s-\gamma}$ and keeping only the lowest order 
corrections, we obtain
\begin{equation}
H_{s-\gamma}=\frac{1}{2}g\mu_B\bm{B}_{\it eff}\cdot\bm{\sigma}+\bm{\mathcal{M}}_{\gamma}\cdot\bm{\sigma}(a^{\dagger}+a)+\hbar\omega 
a^{\dagger}a,\label{effectivehamlong}
\end{equation}
with
\begin{equation}
\bm{B}_{\it eff}=e^{-(l/\lambda_{SO})^2}\bm{B}_{\bm{\eta}\perp}\cdot\bm{\sigma}+B_{\bm{\eta}}\sigma_{\bm{\eta}}
,\label{effZeemanlong}
\end{equation}
\begin{equation}
\bm{\mathcal{M}}_{\gamma}\cdot\bm{\sigma}=eV_0\frac{l}{d}\frac{l}{\lambda_{SO}}\frac{E_Z}{\hbar\omega_0}e^{-(l/\lambda_{SO})^2}(\bm{e}_{\bm{\eta}}\times\bm{l})\cdot\bm{\sigma}.\label{genspinlong}
\end{equation}
We see that the spin-photon interaction is maximal when the magnetic field is perpendicular to 
$\bm{\eta}$, like in the perturbative calculation of the previous section. This  is expected since, 
as in the previous section,  the SOI manifest itself   as an Abelian gauge potential within lowest 
order, although there are two spin-components. For the rest of the paper, we assume now a magnetic 
field perpendicular to  $\bm{\eta}$ so that $B_{\bm{\eta}}=0$, 
$\bm{B}\cdot\bm{\sigma}_{\bm{\eta}\perp}=B\sigma_{\tilde{z}}$ and 
$(\bm{e}_{\bm{\eta}}\times\bm{l})\cdot\bm{\sigma}=\sigma_{\bm{\eta}\perp,\bm{l}}\equiv\sigma_{\tilde{x}}$. Then, the spin-photon Hamiltonian reads
\begin{equation}
H_{s-\gamma}=\frac{1}{2}E_Z^{\it eff}\sigma_{\tilde{z}}+\mathcal{M}_{\gamma}\sigma_{\tilde{x}}(a^{\dagger}+a)+\hbar\omega a^{\dagger}a,\label{partspinlong}\end{equation}
with
\begin{equation}
\mathcal{M}_{\gamma}=eV_0\frac{l}{d}\frac{l}{\lambda_{SO}}\frac{E_Z^{\it eff}}{\hbar\omega_0},\label{partphotlong}
\end{equation}
where $E_Z^{\it eff}=E_Ze^{-(l/\lambda_{SO})^2}$ is the effective Zeeman splitting.

We see that  the SOI  reduces strongly the Zeeman splitting for large values of the ratio 
$l/\lambda_{SO}$. This over-screening of the Zeeman interaction can be understood as follows. After 
performing the transformation (\ref{exacttransform}) there is no SOI present in the system, but the 
magnetic field in the new 'frame' has an oscillatory behavior, as shown in Eq. (\ref{Zeemanx}). 
This means that the magnetic field precesses around  the $x$-direction, the speed of precession 
being given by the strength of the SOI measured through the SO length $\lambda_{SO}$. If the bare 
Zeeman splitting $E_Z$ is much smaller that the orbital level spacing, $E_Z\ll\hbar\omega_0$, the 
electron find itself in the orbital ground state $|0\rangle$ given by $H_0$. Then, if the SOI 
strength is increased, the  precession frequency increases also, so that there are many precessions 
of the magnetic field over small distances. Since this  implies also  small changes of the orbital 
wave-function, this leads  to an average reduction  of the effective Zeeman splitting, as obtained 
above.

\section{Numerical Estimates}   %

We give now some estimates for the coupling $\nu\equiv\mathcal{M}_{ph}^y$ for  QDs in InAs 
nanowires for both geometries shown in Fig. \ref{Fig1}. In the first case, we assume the dots to  
have a  width of $5-10\,{\rm nm}$ ($E_w\approx 10\,{\rm meV}$) and a radius  $R\approx 50\, {\rm 
nm}$ ($\Delta E_0\approx 5\, {\rm meV}$). The electron in the QD is characterized by $m^*=0.023\, 
m_e$, $g\approx\, 2.5$ and $\lambda_{SO}\approx 100\, {\rm nm}$.\cite{Fasth2007} We assume also 
that the 1D cavity is $2$ mm long and $100\,{\rm nm}$ wide, which implies for the fundamental mode 
$\hbar\omega\simeq0.5\,{\rm meV}$ and an rms electric field $E\simeq 100\,{\rm V/m}$.  The Zeeman 
splitting is assumed to be on the same order with the lowest cavity
mode, i.e.  $E_Z^{\it eff}\approx 0.5\, 
{\rm meV}$ ($B\approx 1.75\, {\rm T}$).  Plugging in all the numbers in the formula for $\nu$, Eq. 
(\ref{speffective}) we obtain $\nu\approx 10^{-5}\, {\rm meV}$ which, in the degenerate case 
$E_{Z}^{{\it eff}}=\hbar\omega$, corresponds to a dynamics of the spin-photon system of about $60
{\rm ns}$ (Rabi oscillations between the spin and the cavity). In the second case there is more 
control on the orbital level spacing since the dots are obtained in principle by external gating. 
We now assume a dot radius  $R\approx 10\,{\rm nm}$ ($E_{0t}\simeq 30\,{\rm meV}$), a dot length 
$l\simeq40\,{\rm nm }$ ($\hbar\omega_0\simeq2\,{\rm meV}$) and
$g\approx 10$.\cite{Fuhrer} For $E_Z^{\it eff}\approx0.5 \,{\rm meV}$ we need a
magnetic field $B\approx 0.45\,{\rm T}$. Also, we assume the same lengths for 
the cavity as for the first case so that we obtain $\nu\approx4\cdot10^{-4}\,{\rm meV}$. This gives 
rise to a dynamics of the spin-photon system  of about $2\,{\rm ns}$ in the degenerate limit 
$E_{Z}^{{\it eff}}=\hbar\omega$. We mention that in both cases   the renormalized Zeeman splitting 
is quite different from the bare one, i.e. $E_Z^{\it eff}=0.93 E_Z$ in the first case and 
$E_{Z}^{\it eff}=0.84E_Z$ in the second case.

For the exchange coupling $J$ between two spins one can achieve values as large as $J\approx 10^{-6}\, {\rm meV}$ in the limit of quite small detunings ($\Delta\approx 10^{-4}\, {\rm meV}$)  
for the case in Fig. \ref{Fig1}A,  which eventually translates into a time dynamics of about $500\, 
{\rm ns}$ for coherently swapping the two spins. In the geometry shown in Fig. \ref{Fig1}B the 
exchange coupling $J$ can be much larger, on the order of $J\approx 4\cdot 10^{-5}\,{\rm meV}$ for 
detunigs on the order of $\Delta\approx 4\cdot10^{-3}\,{\rm meV}$, which implies a time dynamics of 
about $20\,{\rm ns}$ for swapping the two spins coherently.

In order  to control the exchange coupling $J$, one should be able in principle to change the 
Zeeman splitting or the orbital level spacing. In InAs QDs the Zeeman splitting can be changed very 
efficiently by changing the dot size along the wire direction,\cite{Fuhrer} in both cases in Fig. 
\ref{Fig1} Considering the case of two QDs in the cavity, one way to decouple  them is by tuning 
the $g$-factors so that $\Delta_1=-\Delta_2$, as  can be seen from  Eq. (\ref{SW}). However, in the 
case of many QDs inside the cavity this will be rather difficult to achieve.

Another possibility is to change  the $g$-factors locally so that the
coupling between the spins reduces due to the reduction of the Zeeman
splitting $E_Z$.  Assuming that a reduction of $J$ by one order  of
magnitude is a good measure for the decoupling, one obtains a
corresponding change in the $g$-factor of the order of $15\%$ in the
first geometry shown in Fig. (\ref{Fig1}). The rather drastic change of $g$-factor was already 
experimentally  demonstrated  for InAs QDs by Bj\"{o}rk $et$ $al.$ [\onlinecite{Fuhrer}]. They  
achieved  a change in the $g$-factor from $|g|=3.5$ to $|g|=2.3$ when the dot size along the 
nanowire was reduced from $10\, {\rm nm}$ to $8\, {\rm nm}$, i.e. a variation of about $30\%$, 
which shows to be sufficient for our scheme in the geometry shown in Fig. \ref{Fig1}. The same can 
be done efficiently for the second geometry, since the dots being gate-defined can be modified 
strongly along the wire axis.

Yet another way to change the exchange coupling $J$ is by changing the orbital confining energy 
$\Delta E_0$. In the first geometry $\nu\sim R^4$, and $J\sim \nu^2$ (assuming two equal 
spin-photon couplings for simplicity) one obtains a dependence $J\sim R^8$. Then, by using top 
gates, for example, one can strongly modify  the exchange coupling $J$ by a small change of the 
orbital energy  $\Delta E_0$. This can be done equally, and maybe more efficiently, for the second 
geometry since, as explained  above, the dots can be modified easily along the wire axis. The 
spin-photon coupling $\nu\sim l^4$, which implies then a scaling of the exchange coupling $J\sim 
l^8$.

\section{Coherent manipulation}%

One way to coherently manipulate and to read-out (measurement) the qubits is by applying an 
external driving field to the cavity with a varying frequency 
$H_{ex}=\epsilon(t)(a^{\dagger}e^{-i\omega_{ex}t}+ae^{i\omega_{ex}t})$,
where $\epsilon(t)$ is the amplitude. In the dispersive limit ($\nu_i/\Delta_i\ll 1$) 
$H_{ex}\rightarrow H_{ex}+[T,H_{ex}]$ so that
\begin{eqnarray}
H_{ex}\simeq\epsilon(t)a^{\dagger}e^{-i\omega_{ex}t}+\sum_{i=1,2}\frac{\nu_i\epsilon(t)}{\Delta_i}\sigma_i^+e^{-i\omega_{ex}t}+{\rm h.c.}\label{EDSR1}
\end{eqnarray}
The control of the $i$-th qubit can now  be realized by tuning the frequency of the driving field 
to $\omega_{ex}=E_{iZ}^{{\it eff}}+\nu_i^2/\Delta_i$, while this condition is not satisfied for the 
other qubits. This gives rise to an electric-dipole spin resonance (EDSR) for the $i$-th qubit, 
similar to that studied by Golovach et al.\cite{Golovach2006} The measurement can be performed by 
tuning the frequency of the driving close to the cavity mode so that we can observe peaks in 
transmission at the positions $\hbar\omega+\sum_{i}(\nu_i^2/\Delta_i)\sigma_z^i$. If  detunings are 
chosen so that all combinations can be distinguished, one can measure all the spins from one shot 
(or at least group of spins).\cite{Blais2004}

A more efficient way to manipulate the spin is to make use of the EDSR-scheme proposed in 
[\onlinecite{Golovach2006}], namely to apply an alternating electric field $\mathcal{E}(t)$ to the 
QD, which, via the electric dipole transitions and the strong SOI, gives rise to an effective 
alternating magnetic  field. Briefly, if only the dipolar coupling to the alternating electric 
field $\mathcal{E}(t)$ is considered, we get $H_{e-el}(t)=e\mathcal{E}(t)y$, with the electric 
field $\mathcal{\bm{E}}(t)$ along $y$-direction. If the system  in Fig. \ref{Fig1}A is considered,  
the effective spin-electric field coupling within first order in SOI becomes 
$H_{s-el}=\langle0|[S,H_{e-el}(t)]|0\rangle \equiv\delta B(t)\sigma_y$, with the fluctuating 
magnetic field $\delta B(t)$ having the form
\begin{equation}
\delta B(t)\sim e\mathcal{E}(t)R\frac{E_Z}{\Delta E_0}\frac{R}{\lambda_{SO}}.\label{EDSR}
\end{equation}
For the case shown in Fig. \ref{Fig1}B we obtain a similar expression for $\delta B(t)$, but with 
the bare Zeeman splitting $E_Z$ substituted with the effective Zeeman splitting $E_Z^{\it eff}$ 
defined after Eq. (\ref{partphotlong}), and the radius $R$ substituted with the dot length $l$. The 
electric field $\mathcal{E}(t)$ is assumed to have an  oscillatory behavior, 
$\mathcal{E}(t)=\mathcal{E}_0\cos{\omega_{ac}t}$ with $\omega_{ac}$ being the frequency of the ac 
electric field. By tuning the frequency of the oscillatory electric field $\omega_{ac}$ in 
resonance with the qubit splitting $E_{Z}^{{\it eff}}$ one can achieve arbitrary rotations of the 
spin on the Bloch sphere on time scales given by the Rabi frequency $\omega_R=\delta 
B(0)/\hbar$.\cite{Golovach2006} We mention that within lowest order in SOI the induced fluctuating 
magnetic field $\bm{\delta{B}}(t)$ is always perpendicular to the applied field $\bm{B}$ and 
reaches the maximum when the applied electric field $\bm{\mathcal{E}}(t)$ points into the same 
direction as $\bm{B}$.\cite{Golovach2006} This is the reason for choosing the electric field along 
the $y$-direction.

We give here also some estimates for the strength of the Rabi frequency $\omega_R$. For this we 
assume the same parameters as in the previous section and we choose for the amplitude of the 
electric field $\mathcal{E}_0\approx 10\, {\rm eV/cm}$. With this values we obtain for the strength 
of the Rabi frequency $\omega_R\approx 10 \,{\rm GHz}$, which gives a time dynamics for the 
electron spin control on the order of $\omega_R^{-1}\approx\, 0.1 {\rm ns}$. This time scale must 
be much shorter than the usual relaxation and decoherence times for the spin in the QD. Finding 
these time scales is the subject of the next section.

\section{Spin relaxation and decoherence}%

We address now the issue of   relaxation and decoherence of the spin in the cavity. There are two 
types of contributions to the relaxation processes,  one arising from the finite decay rate of the 
cavity, $\kappa$, and the other one from the intrinsic relaxation and decoherence  of the spin, 
labeled by $T^{-1}_{1,2}$.
To reach the strong coupling regime described here, the losses must be
smaller than the coupling between the qubits $J$ in the regime of
interest ($\nu^2/\Delta>\kappa,T^{-1}_{1,2}$). Very high-Q factor 1D
electromagnetic cavities were already built  ($Q=\kappa^{-1}\sim
10^{4}-10^{6}$),\cite{Wallraff2004} so that the intrinsic relaxation and decoherence of the qubit 
show up as the limiting factors for reaching the strong coupling regime.

The relaxation and decoherence of the spin-qubit arise mainly from the coupling  to the bath of  
phonons and the collection of nuclei in the QD. The phonon contribution was studied microscopically 
in great detail for the case of gate-defined GaAs QDs in 2DEGs and it was shown that for large 
$B$-fields, similar to  the present case, the main contribution to relaxation comes from the 
deformation potential phonons with a decay time $T_1\sim 10^{-2}-10^{-4}{\rm s}$.\cite{Vitaly} As a 
consequence, a smaller  relaxation time is then expected for InAs QDs since the SOI is one order of 
magnitude larger than in GaAs ($T_1\propto (\lambda_{SO}/R)^2$). However, different from the bulk 
case, the phonon spectrum in nanowires becomes  highly non-trivial due to the mixing of the 
branches by the boundaries,\cite{Norihiko1} leading to a strong modification of the relaxation 
time.

In cylindrical nanowires there are three types of acoustic modes: torsional, dilatational and 
flexural.\cite{book} All these modes couple to the electric charge and, in principle, all of them 
couple also to  the spin for a general SOI Hamiltonian. However, as shown later, this is not 
actually the case for the SOI acting in the two configurations  in Fig. \ref{Fig1}, and only a 
small part of the entire spectrum gives rise to spin relaxation.

As stated above, within  the large Zeeman splitting limit considered in this paper, we can take 
into account only the interaction of the electron with the lattice via the deformation potential. 
The electron-phonon deformation potential interaction is given by 
$H_{e-ph}=\Xi_0\nabla{\bm{u}}(\bm{r},t)$, where $\Xi_0$ is the deformation potential strength and
\begin{equation}
\bm{u}(\bm{r},t)=\frac{1}{\sqrt{N}}\sum_{\bm{k}}[\bm{u}(\bm{k},\bm{r})b_{\bm{k}}(t)+{\rm 
h.c.}],\label{phononfield}
\end{equation}
with the displacement field $\bm{u}(\bm{k},\bm{r})$ given by\cite{book,Norihiko1}
\begin{equation}
\bm{u}(\bm{k},\bm{r})=\nabla\Phi_0+(\nabla\times\bm{e}_z)\Phi_1+(\nabla\times\nabla\times\bm{e}_z)
\Phi_2.\label{displacement}
\end{equation}
The index $\bm{k}\equiv\{q,n,s\}$ quantify the relevant quantum numbers, i.e. the wave-vector along 
the wire, the winding number and the radial number, respectively, $b_{\bm{k}}(t)$ is the 
annihilation operator for phonons, $\bm{e}_z$ is the unit vector along the $z$ direction and
\begin{equation}
\Phi_i=\chi_if_{ns}^i(r)e^{i(n\phi+qz)},\label{phononsolution}
\end{equation}
with $i=0,1,2$, $n=0,\pm1,\pm2\dots$. The functions $f_{ns}^i(r)$ depend only on the 
radius\cite{Norihiko1,Norihiko2} and  $\chi_i$ are normalization factors.
The effective spin-phonon interaction can be found following the same procedure as that used for 
deriving  the spin-photon interaction for both cases in Fig.\ref{Fig1}.

\subsection{Spin-relaxation in strongly-longitudinal confined QDs}

We give here the main steps in the derivation of the relaxation rate
for the case shown in Fig. \ref{Fig1}A. Keeping only terms up to first order in SOI, we obtain
\begin{equation}
H_{s-ph}=\langle0|[S,H_{e-ph}]|0\rangle
,\label{spinphonon}
\end{equation}
with $S$ given in Eq. (\ref{smatrix}) and $|0\rangle$ being the orbital ground-state. Due to the 
circular  symmetry, the first order in SOI term couples  only to the $n=1$ phonons. The resulting 
spin-phonon coupling has the form
\begin{equation}
H_{s-ph}=\frac{1}{2}g\mu_B \delta B_y(t)\sigma_{y}
,\label{spinphonon1}
\end{equation}
with
\begin{equation}
\delta B_y(t)=B\frac{\Xi_0}{\Delta E_0}\frac{R}{\lambda_{SO}}\sum_{q,s}\frac{\mathcal{C}(q,s)}{\sqrt{\mathcal{F}(q,s)\rho_c\,\omega_{q,s}/\hbar}}K_{q,s}^2b^{\dagger}_{\bm{k}}+{\rm h.c.},\label{spinphonon2}
\end{equation}
\begin{equation}
\mathcal{C}(q,s)\approx0.25\int_0^1\frac{dr\,rJ_1(k_{11}r)J_0(k_{10}r)f_{1s}^0(r)}{|J_2(k_{11})J_1
(k_{10})|},\label{overlap}
\end{equation}
where  $K_{q,s}=\omega_{q,s}/c_l$ with $\omega_{q,s}$ being the eigen-modes of the phonon field,
$c_l$ the longitudinal speed of sound in InAs. The normalization function $\mathcal{F}(q,s)$ is 
given by
\begin{equation}
\mathcal{F}(q,s)=\frac{\hbar R^2}{4M\chi_0^2\omega_{\bm{k}}},\label{formfactor}
\end{equation}
where $M$ is the mass of the ions in a unit cell. 

The explicit forms for the  $\omega_{q,s}$ and $\mathcal{F}(\omega_{q,s})$ depend on the boundary 
conditions used for the phonon field. The two quantities  relevant for the boundary conditions are 
the displacement vector $\bm{u}(\bm{r})$ and the stress vector $\bm{t}(r)={\rm T}\bm{e}_r$ at 
$r=R$, with $T$ being the stress tensor\cite{book} and $\bm{e}_r$ being the unit vector along $r$. 
One can now write  $\bm{u}(\bm{r})={\rm \mathcal{U}}\bm{\chi}$ and $\bm{t}(\bm{r})={\rm \mathcal{T}}\bm{\chi}$ with  $\bm{\chi}=(\chi_0,\chi_1,\chi_2)$, where the expressions for  the 
matrices ${\rm\mathcal{U}}$ and ${\rm \mathcal{T}}$ are given in Appendix A.  There are two 
limiting cases for the boundaries. The first case is met when there is zero stress  at the surface, 
i.e. $\bm{t}(R)=0$,\cite{book} with  $\omega_{q,s}$ being the solutions of  $|\mathcal{T}(R)|=0$ 
(free surface boundary conditions or FSBC), while the second limiting case is met when the surface 
is rigid, i.e. $\bm{u}(R)=0$,  with $\omega_{q,s}$ being the solutions of $|\mathcal{U}(R)|=0$ 
(clamped surface boundary conditions or CSBC).  The phonon field is normalized according to the 
following relation\cite{normalization}
\begin{equation}
\frac{1}{\pi R^2}\int_{0}^{2\pi}d\phi\int_0^R dr r\bm{u}^{*}(\bm{k},r,\phi)\cdot\bm{u}(\bm{k},r,\phi)=\frac{\hbar}{2M\omega_{\bm{k}}}.\label{normal
ization}\end{equation}
From the FSBC or CSBC, together with the normalization of the phonon field, one obtains the 
spectrum $\omega_{q,s}$ and the normalization function $\mathcal{F}(q,s)$.

We  now use the effective spin-phonon Hamiltonian with the fluctuating
field given in Eq. (\ref{spinphonon1}) to find the spin relaxation and
decoherence times, $T_1$ and $T_2$, respectively. We mention here that
the fluctuating magnetic field $\delta{B}_y(t)$ is perpendicular to
the external one $\bm{B}$ such that there is no pure dephasing coming
from the interaction of the spin with phonons in lowest order in
SOI. In fact, as shown previously,\cite{Vitaly} this is valid for any
type of baths, be it phonons, particle-hole excitations etc. 

In the following we derive the expressions of the $T_1$ and $T_2$ times resulting from the 
fluctuating field $\delta B_y(t)$. For this
we need to compute the bath correlator
\begin{equation}
J_{yy}(\omega)=\left(\frac{g\mu_B}{2\hbar}\right)^2\int_{0}^{\infty}dte^{-i\omega 
t}<\delta{B_y(0)}\delta{B_y(t)}>,\label{correlator}
\end{equation}
where the brackets $<...>$  means  tracing over the phonon bath being at thermal equilibrium at 
temperature T. The relaxation time within the Bloch-Redfield approach is given in the present
particular case (the $B$-field along $x$-direction) by (see Ref. \onlinecite{Vitaly,Massoud})
\begin{equation}
T_1^{-1}={\rm Re}(J_{yy}(\omega_Z^{{\it eff}})+J_{yy}(-\omega_Z^{{\it eff}})),\label{timerelax1}
\end{equation}
 with $\omega_Z^{{\it eff}}=E_Z^{{\it eff}}/\hbar$.
Making use of Eq. (\ref{correlator}) we then finally obtain for the relaxation
rate
\begin{equation}
T_1^{-1}=T_{(0)1}^{-1}\left(\frac{\omega_Z^{{\it eff}}R}{
c_l}\right)^5\sum_s\left(\left|\frac{\partial q}{\partial
\omega_{q,s}}\right|\frac{\mathcal{C}^2(q,s)}{\mathcal{F}(q,s)}\right)_{\omega_{q,s}\equiv\omega_Z
^{{\it eff}}},\label{timerelax2}
\end{equation}
where
\begin{equation}
T_{(0)1}^{-1}\approx0.05\frac{\delta^2\hbar}{\rho_cR^5}\left(\frac{\Xi_0}{\Delta 
E_0}\right)^2\left(\frac{R}{\lambda_{SO}}\right)^2.\label{time0}
\end{equation}
In the above expression  $\delta=E_Z/E_Z^{{\it eff}}$,  and the functions $\mathcal{C}(q,s)$ and 
$\mathcal{F}(q,s)$ are defined in
Eqs. (\ref{overlap},\ref{formfactor}). We mention that within first
order in SOI the decoherence time $T_2$ induced by phonons satisfies
$T_2=2T_1$ since, as mentioned before, the fluctuating magnetic field
induced by phonons $\bm{\delta{B}}$ is perpendicular to the applied one $\bm{B}$.
\begin{figure}[t]
\scalebox{0.31}{\includegraphics{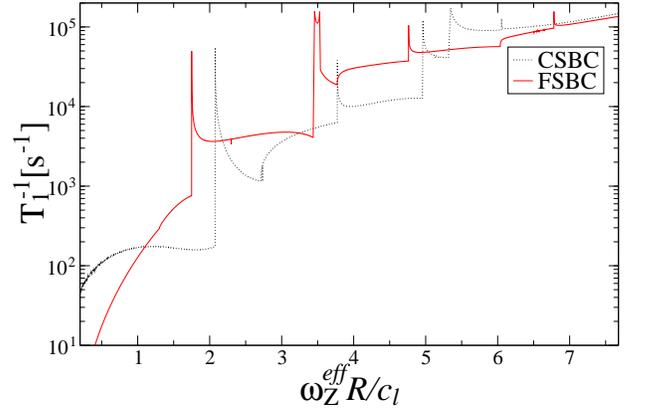}}
\caption{The relaxation rate $T_1^{-1}$ as a function of the ratio $\omega_Z^{{\it eff}}R/c_l$, for 
both FSCB and CSBC (see text for explanations of FSBC and CSBC). Here  $\hbar c_l/R\simeq 
0.6\cdot10^{-4}\,{\rm eV}$ ($c_l\simeq 4\cdot 10^3 \,{\rm m/s}$ and $R\simeq 50 \,{\rm nm}$) 
corresponding to a magnetic field $B\simeq 0.2 \,{\rm T}$, for $g=2.5$.}
\label{FigRelax}
\end{figure}
In Fig. \ref{FigRelax} we plot the relaxation time as a function of
the ratio $\omega_Z^{{\it eff}}R/ c_l$, for $R=50\,{\rm nm}$ and
$c_l=4\cdot 10^3\,{\rm m/s}$.  We see that the relaxation rate
exhibits peaks as a function of the effective Zeeman  splitting
$E_Z^{\it eff}$. This is due to the finite size in the transverse
direction which gives rise to phonon branches. Each new peak appears
when $E_{Z}^{\it eff}$ reaches a new energetically  higher
branch.  Note that although the relaxation rate  seems to diverge when reaching a new peak, in reality this does not happen since there are many processes which broaden the phonon DOS at these special points, like phonon-phonon scattering, phonon-substrate scattering, etc. The usual branch splitting is on the order of
$\omega_{ph}^R\equiv c_l/R $, which  stands for the phonon frequency
in bulk material with the wave-length equal to the dot size $R$. This
frequency $\omega_{ph}^R$ (or energy, when expressed as
$\hbar\omega_{ph}^R$) is the parameter which characterizes the
dominant mechanism for the phonon-induced spin relaxation, which can
be due to piezoelectric-potential  or deformation-potential
phonons. In the limit $\omega^{{\it eff}}_Z\ll \omega_{ph}^R$ the
piezo-phonons give the main contribution to the relaxation rate
$T_1^{-1}$, while in the opposite case, $\omega^{\it eff}_Z\gg
\omega_{ph}^R$, the main contribution to the relaxation rate
$T_1^{-1}$ is given  by deformation-potential phonons.\cite{Vitaly}
Here we are in neither of the two limits, but in the range where
Zeeman splitting is slightly larger than $\hbar\omega_{ph}^R$,
i.e. $\omega^{\it eff}_Z\geq\omega_{ph}^R$. However,  taking into
account only the deformation potential mechanism  should give the
right order of magnitude for the relaxation rate. We mention here that
the relaxation rate $T_1^{-1}$ in the low energy limit ($\omega_Z^{\it
  eff}R/c_l<1$) is given predominantly by the longitudinal linear  in $q$ mode ($\omega_{\it
long}(q)=c_lq$) and the bending mode, square in $q$ ($\omega_{\it bend}(q)=Bq^2$, with $B$ being a constant which depends on $R$).\cite{book}
 
We see from Fig. \ref{FigRelax} that each new phonon branch gives a strong enhancement of the 
relaxation rate $T_1^{-1}$, since it adds more phonon density of states. However, we see also that 
before the first peak, i.e. before reaching the first  new branch,  there is  little spin 
relaxation ($T_1\leq10^{-3} {\rm s}$) for both FSBC and CSBC. This energy scale corresponds to a 
Zeeman splitting  $E_{Z}^{{\it eff}}\approx 10^{-4} {\rm eV}$($E_Z^{{\it eff}}\approx1.2\cdot10^{-4} {\rm eV}$) for FSBC (CSBC). 

If one tunes the ${\it effective}$ Zeeman splitting $E_{Z}^{{\it eff}}$ below the first peak, the 
relaxation rate of the qubit becomes very small, and the fact that $E_Z^{{\it eff}}$ and {\it not} 
$E_{Z}$ has to be tuned  is practically an advantage for reasonably strong SOI since we need quite 
large $E_Z$ for having large $g\propto E_Z$. In the present case $E_Z^{{\it eff}}/E_Z\approx 0.93$, 
and for larger SOI  this ratio will  be even smaller.

\subsection{Spin relaxation in strongly-transverse confined QDs}

We give here a brief description  of the phonon-induced spin relaxation for the case shown in Fig. 
\ref{Fig1}B. We first mention that due to the strong confinement in the transverse direction we can 
average the electron-phonon interaction over the transverse orbital ground state $|0_t\rangle$. 
Since for the ground state wave function we have $m=0$ (see Eq. (\ref{wavefunction})), the only 
modes which couple to the electron, and thus eventually to the spin, are the $n=0$ modes of the 
phonon field in Eq. (\ref{phononfield}). Then, the problem of relaxation simplifies considerably.

\begin{figure}[t]
\scalebox{0.31}{\includegraphics{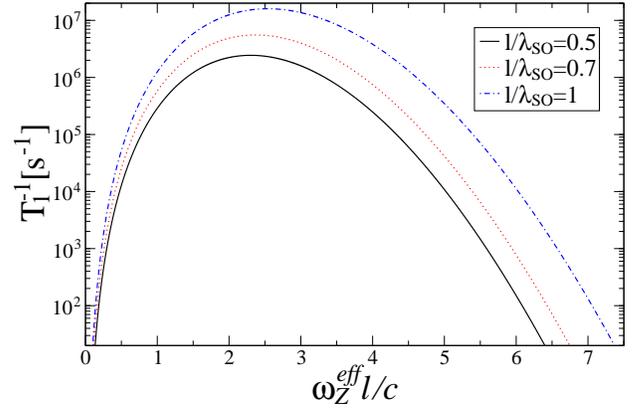}}
\caption{The relaxation rate $T_1^{-1}$ as a function of the ratio $\omega_Z^{{\it eff}}l/c$ for 
three different ratios $l/\lambda_{SO}$ and with FSBC (see text).}
\label{FigRelaxLong}
\end{figure}

The transformation $H_{e-ph}\rightarrow e^{-S}H_{e-ph}e^S$, with  $S$ given in Eq. 
(\ref{exacttransform}), although exact, does not lead  to a coupling of the spin to the phonon 
field since both the  electron-phonon interaction Hamiltonian $H_{e-ph}$ and $S$ contain only 
coordinate $x$ operator, i.e. they commute. After this transformation, however, we are left with no 
SOI term, but with the  $x$-dependent Zeeman coupling in Eq. (\ref{Zeemanx}). We now perform a 
second transformation $H_{e-ph}\rightarrow e^{-S'}H_{e-ph}e^{S'}$ with $S'$ given  before Eq. 
(\ref{effectivehamlong}), under the assumption $E_Z\ll\hbar\omega_0$. Then, within first order in 
$E_Z/\hbar\omega_0$  we obtain for the spin-phonon Hamiltonian $H_{s-ph}$ the following expression
\begin{equation}
H_{s-ph}=\langle0|[S',H_{e-ph}]|0\rangle,\label{sphononlong}
\end{equation}
where we averaged also over the ground-state $|0\rangle$ of the orbital Hamiltonian $H_0$. The 
spin-phonon Hamiltonian $H_{s-ph}$ reads
\begin{equation}
H_{s-ph}=\frac{1}{2}g\mu_B\delta B_{\tilde{x}}(t)\sigma_{\tilde{x}}+\frac{1}{2}g\mu_B \delta 
B_{\tilde{z}}(t)\sigma_{\tilde{z}},\label{splong}
\end{equation}
with
\begin{equation}
\delta B_{\tilde{x},\tilde{z}}(t)=B_{\it eff}\frac{\Xi_0}{\hbar\omega_0}\sum_{q,s}\frac{M_{s-ph}^{\tilde{x},\tilde{z}}(q)}{\sqrt{2\mathcal{F}(q,s)\rho_c\omega_{q,s}/\hbar}}K^2_{\bm{k}}b_{\bm{k}}^{\dagger}+{\rm h.c.}, 
\label{fluctuatinglong}
\end{equation}
and $\bm{k}\equiv\{q,s\}$. The functions $M_{s-ph}^{\tilde{x},\tilde{z}}$ are given by the 
following expressions
\begin{equation}
M_{s-ph}^{\tilde{x}}(q)={\rm SinhInt}\left(\frac{l^2q}{\lambda_{SO}}\right)\label{xcoupling}
\end{equation}
\begin{equation}
M_{s-ph}^{\tilde{z}}(q)=\gamma-{\rm CoshInt}\left(\frac{l^2q}{\lambda_{SO}}\right)+{\rm Log}\left(\frac{l^2q}{\lambda_{SO}}\right),\label{zcoupling}
\end{equation}
where $\gamma=0.577$ is  the Euler constant, Log($x$) is   the natural logarithm, while  the 
special functions SinhInt($x$) and CoshInt($x$) are defined as\begin{equation}
{\rm SinhInt}(x)=\int_0^xdt \frac{\sinh{(t)}}{t}\label{special1}
\end{equation}
\begin{equation}
{\rm CoshInt}(x)=\gamma+{\rm Log}(x)+\int_0^xdt \frac{\cosh{(t)}-1}{t}.\label{special2}
\end{equation}
We see that,  there is both relaxation
and pure dephasing of the spin due to spin-phonon
interaction. However, since the deformation-potential phonons is
superohmic (even in  1D case for deformation-potential phonons), the pure dephasing rate 
vanishes\cite{Weiss} so that we  retain in the following only the first term in Eq. (\ref{splong}). 
The relaxation rate $T_1^{-1}$ can be found by the same procedure as in the previous case and reads
\begin{equation}
T_{1}^{-1}={\rm Re}(J_{\tilde{x}\tilde{x}}(\omega_Z^{\it eff})+J_{\tilde{x}\tilde{x}}(-\omega_Z^{\it eff})),\label{timelong}
\end{equation}
where the correlation function $J_{\tilde{x}\tilde{x}}$ is defined in Eq. (\ref{correlator}) with 
$y\rightarrow\tilde{x}$, and $\omega_Z^{\it eff}=E_Z^{\it eff}/\hbar$, as before. The expression 
for the relaxation rate $T_1^{-1}$ becomes
\begin{equation}
T_1^{-1}=T_{(0)1}^{-1}\left(\frac{\omega_Z^{\it eff}l}{
c_l}\right)^5\sum_s\left(\bigg|\frac{\partial q}{\partial\omega_{q,s}}\bigg|\frac{\widetilde{\mathcal{M}}_{s-ph}^{2\tilde{x}}(q)}{\mathcal{F}(q,s)}\right)_
{\omega_{q,s}=\omega_Z^{\it eff}},\label{timelong}
\end{equation}
 where
\begin{equation}
T_{(0)1}^{-1}=\frac{\hbar}{2\pi \rho_c R^2l^3}\left(\frac{\Xi_0}{\hbar\omega_0}\right)^2\label{timeconstlong}
\end{equation}
and
\begin{equation}
\widetilde{\mathcal{M}}_{s-ph}^{\tilde{x}}(q)=\mathcal{M}_{s-ph}^{\tilde{x}}(q)e^{-q^2l^2/8}.\label{suppression}
\end{equation}
In order to find now the dependence of the relaxation rate $T_{1}^{-1}$ on the effective Zeeman 
splitting $\omega_Z^{\it eff}$, we have to find first the phonon eigen-frequencies $\omega_{q,s}$. 
This can be done following the  same steps as  in the previous section, depending which kind of 
boundary conditions are used, i.e. FSBC or CSBC. As mentioned earlier, the average distance between 
the branches $s$ is on the order of $\omega_{ph}^R=c_l/R$. Then, since  $R\ll l$, and also due to 
the gaussian suppression  in Eq. (\ref{suppression}), it is enough to consider  in Eq. 
(\ref{timelong}) only the lower branch $s=1$. If we now  assume FSBC and the limit $qR\ll1$, the 
phonon eigen-frequency becomes linear in $q$, i.e. $\omega_{q,1}\equiv\omega(q)=cq$, 
with\cite{book}
\begin{equation}
c=c_t\sqrt{\frac{3c_l^2-4c_t^2}{c_l^2-c_t^2}}.\label{speed}
\end{equation}
The  normalization function $\chi_0$ acquires also a  simple form  in this limit, and reads
\begin{equation}
\chi_0=\frac{c_l^2}{3c_l^2-4c_t^2}\frac{R}{q}\sqrt{\frac{\hbar}{2M\,c\,q}}.\label{normalizationlon
g}
\end{equation}
After inserting  in Eq. (\ref{timelong}) the expressions for $\omega(q)$  and $\chi_0$, we obtain 
for the relaxation rate $T_1^{-1}$  (FSBC)  the final expression
\begin{equation}
T_{1}^{-1}=\frac{T_{(0)1}^{-1}}{2}\left(\frac{c^2}{3c_l^2-4c_t^2}\right)^2\left(\frac{\omega_Z^{\it eff}l}{c}\right)^3\widetilde{\mathcal{M}}_{s-ph}^{\tilde{x}2}(\omega_Z^{\it eff}l/c).\label{finalrelaxlong}
\end{equation}
In Fig. \ref{FigRelaxLong} we plot the relaxation rate $T_{1}^{-1}$ as a function of the 
dimensionless parameter $\omega_Z^{\it eff}l/c$ for different SOI strengths measured through the 
ratio $l/\lambda_{SO}$. We assumed here  $R=10\, {\rm nm}$ and  $l=50\, {\rm nm}$, which gives  
$\hbar c/l\equiv\hbar\omega_{ph}^{l}=0.05\,{\rm meV}$ and $\hbar c_l/R\equiv 
\hbar\omega_{ph}^{R}=0.25\,{\rm meV}$.  We see in Fig. \ref{FigRelaxLong} that the relaxation rate 
$T_1^{-1}$ is quite large ($T_1^{-1}\sim 10^{5}-10^{7}\,{\rm s^{-1}}$) for $\omega_Z^{\it eff}/\omega_{ph}^l\sim 1-5$, i.e. when these energies are comparable. However, there is need for a 
large effective Zeeman splitting $E_Z^{\it eff}\gg\hbar\omega_{ph}^l$ to achieve a large 
spin-photon coupling $\mathcal{M}_{\gamma}$. 
At the same time, one should stay still below the next phonon
branch since above it we find a substantial increase for the
relaxation rate. Since this next phonon branch lies somewhere around
$2\hbar\omega_{ph}^R\approx 0.5\,{\rm meV}$, the condition for strong
spin-phonon coupling and weak relaxation becomes
$\hbar\omega_{ph}^l\ll E_Z^{\it eff}<2\hbar\omega_{ph}^R$. In this
regime we are actually satisfying also the necessary condition that
$E_Z/\hbar\omega_0\ll1$, since 
for $l=50\,{\rm nm}$ we have $\hbar\omega_0=1.3 \, {\rm meV}$.  We mention that for CSBC the  
phonon spectrum is gaped, and, in consequence,  there is no phonon-induced relaxation of the spin for 
Zeeman splittings $E_Z^{\it eff}$ smaller than this gap $\Delta_{ph}$. This energy (gap) is on the 
order of $\Delta_{ph}\sim 2\hbar\omega_{ph}^R=0.5\,{\rm meV}$. Note
the non-monotonic behavior of the relaxation rate as a function of the
effective Zeeman splitting (see Fig. \ref{FigRelaxLong}). This
non-monotonicity has the same origin as in GaAs QDs,\cite{Vitaly} and
comes from the fact that for increasing Zeeman splitting the
wave-length of the phonon decreases, and when this becomes less than
the dot length the phonons decouple from the electron (i.e. the
electron-phonon coupling averages to zero). A similar non-monotonic
effect has been recently observed in GaAs double QDs.\cite{Meunier} 

The spin decoherence time due to phonon processes is given by $T_2 = 2T_1$ such that
the main source for decoherence comes from the hyperfine interaction
between the electron and the surrounding nuclei.
This time scale, $T_2^*$, is given by\cite{Khaetskii2003,Coish2004}
\begin{equation}
T_2^*=\frac{N}{\sqrt{A}},\label{decoherence}
\end{equation}
where $N$ is the number of nuclei in the sample and $A$ is the
hyperfine constant. We see that the larger the number of nuclei,
i.e. the bigger the dot, the  longer is the pure decoherence time
$T_{2}^*$ for the electron.  In a typical GaAs QDs ($R=30\,{\rm nm}$
and $l=5\,{\rm nm}$) this time scale  is on the order of $T_2^*\sim
10^{-8}\, {\rm s}$,\cite{Khaetskii2003,Coish2004} which corresponds
also to our two cases described in the paper. However, again
like in GaAs, we expect that coherently driving the qubit will
prolong the $T_2^*$ time up to $10^{-6}\,{\rm s}$ and  with echo up
to $10^{-5}\,{\rm s}$.\cite{Petta2005} Moreover, like in GaAs QDs,
one can make use of state narrowing procedures,\cite{Klauser,
Stepanenko} which should lead to a further substantial enhancement
of $T_2^*$ due to nuclear spins, and possibly reach the SOI induced
limit of $10^{-1}-10^{-4}\,{\rm s}$ calculated above for large
magnetic field strengths.


\section{Conclusions}%

We have proposed and studied an  efficient way to implement  spin
qubits localized in InAs nanowires coupled to a 1D electromagnetic
transmission line (cavity) via  SOI. We have analyzed two
experimentally achievable configurations of the system. In the first
case the electronic confinement is much stronger along the nanowire
axis than in the transverse direction (large-radius nanowires), while
the other case corresponds to the opposite limit (small-radius
nanowires).  We have found  a reasonably strong coupling between the spin
and the cavity modes due to both strong vacuum fluctuations in the
cavity  and strong SOI in InAs. We  also have shown that this
spin-photon coupling can allow  for coupling between two (or several) distant spins,
depending on the detuning of the Zeeman splittings $E_{iZ}^{\it eff}$
from the cavity mode $\hbar\omega$. The SOI-induced exchange coupling
$J$ between two spins can be controlled  by  electrical fields only,
e.g. by changing the $g$-factor and/or orbital level spacing. Also,
single-spin rotations can be performed efficiently by electric fields
only, through the EDSR mechanism. In principle, the price one has to
pay in  strong SOI materials  is strong coupling to the charge
environment  which then relaxes and decoheres the spin. However, we
have studied the relaxation of the spin due to the lattice vibrations
in the InAs nanowires for both configurations, and shown that the time scale
for the spin-decay is on the order of milliseconds for strong magnetic
fields ($B\sim 0.5-1\, {\rm T}$), much larger than the times
associated with the spin-photon dynamics, which takes place on times
scales on the  order of
$10^{-8}-10^{-7}\,{\rm s}$. This fact is due to the quasi 1D structure of the system where the 
phonon spectrum shows discrete branches,  very different from  the bulk limit.

We stress here also  that the coupling of the quantized modes of the transmission line to the spin 
degree of freedom via SOI is not restricted to QDs in semiconductor nanostructures. In principle, 
this coupling should be possible in  other  spin-orbit coupled systems too, like nitrogen-vacancy 
centers (NV-centers),\cite{Lukin2006,HansonNV2006} molecular 
magnets,\cite{Loss2001,Tejada2001,Lehmann2007} magnetic nanorings,\cite{Filippo2006} etc. In these 
systems there is usually a strong zero-field splitting (ZFS) of the lowest spin-multiplet 
attributed to  SOI or to dipole-dipole interaction. This would  allow for an efficient coupling of 
the  electric fields, quantum or classical, to the spin degree of freedom and finally providing a 
mechanism  for an all-electrical implementation of  spin-based quantum information processing.

As a final remark, we mention that the present scheme can be also used  to form hybrid structures 
where spin-qubits are integrated together with other types of qubits in the same 1D transmission 
line. For example, one can envision a setup where a spin-qubit is coupled via the cavity modes to 
superconducting qubit as the one studied in Ref.~\onlinecite{Blais2004} so that one can transfer  
arbitrary states between the two qubit-systems.

\section{Acknowledgments}

We thank  D. Bulaev  and G. Burkard for useful discussions.
We acknowledge financial support from the Swiss NSF, the NCCR Nanoscience Basel, and JST ICORP.
\\
\\

\appendix%

\section{Displacement and stress tensor for cylindrical nanowires}%
In this Appendix we give explicit formulas for the displacement $\bm{u}(\bm{r})$ and stress 
$\bm{t}(\bm{r})$ vectors, respectively. We can write the displacement vector 
$\bm{u}(\bm{r})=(u_r,u_{\phi},u_z)$ from Eq. (\ref{displacement}) in components
\begin{equation}
u_{i}(\bm{r},t)=\sum_{j}U_{ij}(r)\bm{\chi}_je^{i(n\phi+qz-\omega t)},\;\;\;i=r,\phi,z,
\end{equation}
with $\bm{\chi}_i=(\chi_0,\chi_1,\chi_2)$ and the matrix $\mathcal{U}(r)$ having the form
\begin{equation}
\mathcal{U}(r)=\left(
\begin{array}{ccc}
\frac{\partial}{\partial r}f_{0n}(r) & i\frac{n}{r}f_{1n}(r) & iq\frac{\partial}{\partial 
r}f_{2n}(r) \\i\frac{n}{r}f_{0n}(r) & -\frac{\partial}{\partial r}f_{1n}(r) & 
-\frac{nq}{r}f_{2n}(r)\\iqf_{0n}(r) & 0 & k_1^2f_{2n}(r)
\end{array}
\right).
\end{equation}
The other relevant quantity for the elastic problem is the stress tensor $T$.\cite{book} In order 
to obtain $T$, we first have  to find the strain tensor $S$ as a function of displacement 
$\bm{u}(\bm{r})$. The independent components of the strain tensor  coordinates have expressions (in 
cylindrical coordinates)\cite{book} of the form
\begin{eqnarray}
S_{rr}&=&\frac{\partial u_r}{\partial r}\nonumber\\S_{\phi\phi}&=&\frac{1}{r}\left(\frac{\partial 
u_{\phi}}{\partial \phi}+u_r\right)\nonumber\\S_{zz}&=&\frac{\partial u_z}{\partial 
z}\nonumber\\S_{r\phi}&=&\frac{1}{2r}\left(\frac{\partial u_r}{\partial 
\phi}+r^2\frac{\partial}{\partial 
r}\left(\frac{u_r}{r}\right)\right)\nonumber\\S_{z\phi}&=&\frac{1}{r}\frac{\partial u_z}{\partial 
\phi}+\frac{\partial u_{\phi}}{\partial z}\nonumber\\S_{rz}&=&\frac{1}{2}\left(\frac{\partial 
u_r}{\partial z}+\frac{\partial u_z}{\partial r}\right).
\end{eqnarray}
The stress tensor, $T$ , which quantifies the surface forces, is related to the strain tensor $S$ 
by the elastic modulus constants.\cite{book} Since we are interested in the boundary conditions at 
the surface of the cylinder, the relevant part of the stress tensor is given by the stress vector 
$\bm{t}=T\bm{e}_r$, with $\bm{e}_r$ being the unit vector along the radius. We write here only 
these relevant parts of the stress tensor $T$ as a function of the strain tensor components
\begin{eqnarray}
T_{rr}&=&\rho(c_l^2-2c_t^2)(S_{rr}+S_{\phi\phi}+S_{zz})+2\rho 
c_t^2S_{rr}\nonumber\\T_{r\phi}&=&2\rho c_t^2S_{r\phi}\nonumber\\T_{rz}&=&2\rho c_t^2S_{rz},
\end{eqnarray}
and $\bm{t}=(T_{rr},T_{r\phi},T_{rz})$.
 We write now the relevant stress vector $\bm{t}$, which is given explicitly by the following 
relation
\begin{widetext}
\begin{equation}
\left(\begin{array}{c}
T_{rr}\\T_{r\phi}\\T_{rz}
\end{array}
\right)=\rho\left(
\begin{array}{ccc}
c_l^2\frac{\partial}{\partial r}+(c_l^2-2c_t^2)\frac{1}{r} &
(c_l^2-2c_t^2)\frac{1}{r}\frac{\partial}{\partial \phi} & (c_l^2-2c_t^2)\frac{\partial}{\partial
z}\\
c_t^2\frac{1}{r}\frac{\partial}{\partial \phi} & c_t^2(\frac{\partial}{\partial r}-\frac{1}{r}) &
0\\
c_t^2\frac{\partial}{\partial z} & 0 & c_t^2\frac{\partial}{\partial r}
\end{array}\right)
\left(\begin{array}{c}u_r\\u_{\phi}\\u_z\end{array}\right).\label{displacement1}
\end{equation}
\end{widetext}
We can bring the stress matrix to the same form as we did for the displacement, namely 
$t_{i}(\bm{r})=\sum_{j}\mathcal{T}_{ij}(r)\bm{\chi}_je^{i(n\phi+qz-\omega t)}$, with the matrix 
$\mathcal{T}$ having the explicit form
\begin{widetext}
\begin{equation}
\mathcal{T}(r)=\left(
\begin{array}{ccc}
\left(2c_t^2\frac{\partial^2}{\partial 
r^2}-(c_l^2-2c_t^2)\left(\frac{\omega}{c_l}\right)^2\right)f_{0n} & 
2inc_t^2\frac{\partial}{\partial r}\left(\frac{f_{1n}}{r}\right) & 
2iqc_t^2\frac{\partial^2}{\partial r^2}f_{2n}\\  2inc_t^2\frac{\partial}{\partial 
r}\left(\frac{f_{0n}}{r}\right)  & -c_t^2\left(2\frac{\partial^2}{\partial r^2}+k_1^2\right)f_{1n} 
&  -2qnc_t^2\frac{\partial}{\partial r}\left(\frac{f_{2n}}{r}\right) \\ 
2ic_t^2q\frac{\partial}{\partial r}f_{0n} & -c_t^2\frac{nq}{r}f_{1n} & 
c_t^2(k_1^2-q^2)\frac{\partial}{\partial r}f_{2n}.\label{stress}
\end{array}
\right).
\end{equation}
\end{widetext}


\begin{references}

\bibitem{LDV}
D. Loss and D. P. DiVincenzo, Phys. Rev. A {\bf 57}, 120 (1998).
\bibitem{Cerletti}
V. Cerletti, W. A. Coish, O. Gywat, and D. Loss,
Nanotechnology {\bf 16}, R27 (2005).
\bibitem{Hanson2007}
R. Hanson, L. P. Kouwenhoven, J. R. Petta, S. Tarucha, and L. M. K. Vandersypen,
arXiv:cond-mat/0610433.
\bibitem{NielsenChuang}
M. A. Nielsen and I. L. Chuang,
{\em Quantum Computation and Quantum Information}
(Cambridge U. Press, NewYork, 2000).
\bibitem{Fasth2007}
C. Fasth, A. Fuhrer, L. Samuelson, V. N. Golovach, and D. Loss,
Phys. Rev. Lett. {\bf 98}, 266801 (2007).
\bibitem{Fuhrer}
M. T. Bj\"{o}rk, A. Fuhrer, A. E. Hansen, M. W. Larsson, L. E. Froberg, and L. Samuelson,
Phys. Rev. B {\bf 72}, 201307({\bf R}) (2005).
\bibitem{KAT}
L.P. Kouwenhoven, D. G. Austing, and S. Tarucha,
Rep. Prog. Phys. {\bf 64} 701 (2001).
\bibitem{Tarucha1996}
S. Tarucha, D. G.  Austing, T.  Honda, R. J.  van der Hage, and L. P. Kouwenhoven,
Phys. Rev. Lett. {\bf 77}, 3613 (1996).
\bibitem{Ciorga2000}
M. Ciorga, A. S. Sachrajda, P.  Hawrylak, C. Gould, P.  Zawadzki, S. Jullian, Y.  Feng, and Z.  
Wasilewski,
Phys. Rev. B {\bf 61}, 16315 (2000).
\bibitem{Elzerman2003}
J. M. Elzerman, R.  Hanson, J. S.  Greidanus, L. H.  Willems van Beveren, S. de
Franceschi, L. M.  Vandersypen, S.  Tarucha, and L. P.  Kouwenhoven,
Phys. Rev. B {\bf 67}, 161308 (2003).
\bibitem{Hayashi2003}
T. Hayashi, T.  Fujisawa, H. D.  Cheong, Y. H.  Jeong,  and Y. Hirayama,
Phys. Rev. Lett. {\bf 91}, 226804 (2003).
\bibitem{Petta2004}
J. R. Petta, A. C.  Johnson, C. M.  Marcus, M. P.  Hanson, and A. C.  Gossard,
Phys. Rev. Lett. {\bf 93}, 186802 (2004).
\bibitem{Elzerman2004}
J. M. Elzerman,R.  Hanson, L. H.  Willems van Beveren, B.  Witkamp, L. M. K.  Vandersypen, and L. 
P.  Kouwenhoven,
Nature, {\bf 430}, 431 (2004).
\bibitem{Amasha2006}
S. Amasha, K. MacLean, I. Radu, D. M. Zumbuhl,
M. A. Kastner, M. P. Hanson, and A. C. Gossard,
arXiv:cond-mat/0607110 (2006).
\bibitem{JohnsonPettaT2Nature}
A.C. Johnson, J.R. Petta, J.M. Taylor, A. Yacoby, M.D. Lukin, C.M. Marcus, M.P. Hanson, and A.C. 
Gossard,
Nature {\bf 435}, 925 (2005).
\bibitem{Petta2005}
J. R. Petta, A. C.  Johnson, J. M.  Taylor, E. A.  Laird, A.  Yacoby, M. D.
Lukin, C. M.  Marcus, M. P.  Hanson, and A. C.  Gossard,
Science, {\bf 309}, 2180 (2005).
\bibitem{Engel}
H.-A. Engel and D. Loss,
Phys. Rev. B {\bf 65}, 195321 (2002).
\bibitem{Koppens2006}
F. H. L. Koppens, C. Buizert, K. J. Tielrooij, I. T. Vink,
K. C. Nowack, T. Meunier, L. P. Kouwenhoven, and
L. M. K. Vandersypen,
Nature {\bf 442}, 766 (2006).
\bibitem{Nowack}
K. C. Nowack, F. H. L. Koppens, Yu. V. Nazarov, and L. M. K. Vandersypen, 
arXiv:0707.3080.
\bibitem{Golovach2006}
V. N. Golovach,  M. Borhani, and D. Loss,
Phys. Rev. B {\bf 74}, 165319 (2006).
\bibitem{Laird07}
E. A. Laird, C. Barthel, E. I. Rashba, C. M. Marcus, M. P. Hanson, and A. C. Gossard, 
arXiv:0707.0557. 
\bibitem{Bjork2004}
M. T. Bj\"ork, C. Thelander, A. E. Hansen, L. E. Jensen, M. W. Larsson, L. R. Wallenberg, and L. 
Samuelson,
Nano Lett. {\bf 4}, 1621 (2004).
\bibitem{Fasth2005}
C. Fasth, A. Fuhrer, M. T. Bj\"ork, and L. Samuelson,
Nano Lett. {\bf 5}, 1487 (2005).
\bibitem{Pfundt2006}
A. Pfund, I. Shorubalko, R. Leturcq, and K. Ensslin,
Appl. Phys. Lett. {\bf 89}, 252106 (2006).
\bibitem{Shorubalko2007}
I. Shorubalko, A. Pfund, R. Leturcq, M. T. Borgstr\"om, F. Gramm, E. M\"uller, E. Gini, and  K.
Ensslin,
Nanotechnology {\bf 18}, 044014 (2007).
\bibitem{Rashba2003}
E. I. Rashba and Al. L. Efros,
Phys. Rev. Lett. {\bf 91}, 126405 (2003).
\bibitem{Kato2004}
Y. Kato, R. C. Myers, A. C. Gossard, and D. D. Awschalom,
Nature (London) {\bf 427}, 50 (2004).
\bibitem{Duckheim2006}
M. Duckheim and D. Loss,
Nat. Phys. {\bf 2}, 195 (2006).
\bibitem{Trif2007}
M. Trif, V. N. Golovach, and D. Loss,
Phys. Rev. B {\bf 75}, 085307 (2007).
\bibitem{Terhal}
K. M. Svore, B. M. Terhal, and D. P. DiVincenzo, 
Phys. Rev. A {\bf 72}, 022317(2005).
\bibitem{Imamoglu}
A. Imamo\=glu, D. D. Awschalom, G. Burkard, D. P. DiVincenzo, D. Loss, M. Sherwin, and A. Small,
Phys. Rev. Lett. {\bf 83}, 4204 (1999).
\bibitem{Berezovsky2006}
J. Berezovsky, M. H. Mikkelsen, O. Gywat, N. G. Stoltz, L. A. Coldren, and D. D. Awschalom,
Science {\bf 314}, 1916 (2006).
\bibitem{Atature2007}
M. Atature, J. Dreiser, A. Badolato, and A. Imamoglu,
Nat. Phys.  {\bf 3}, 101 (2007).
\bibitem{Wallraff2004}
A. Wallraff, D. I. Schuster, A. Blais, L. Frunzio, R.-S. Huang,
J. Majer, S. Kumar, S. M. Girvin, and R. J. Schoelkopf,
Nature (London) {\bf 431}, 162 (2004).
\bibitem{Blais2004}
A. Blais, R.-S. Huang, A. Wallraff, S. M. Girvin, and R. J.
Schoelkopf, 
Phys. Rev. A {\bf 69}, 062320 (2004).
\bibitem{Gywat}
O. Gywat, F. Meier, D. Loss, and D. D. Awschalom,
Phys. Rev. B {\bf 73}, 125336 (2006).
\bibitem{Lukin2004}
L. Childress, A. S. S\o rensen, and M. D. Lukin,
Phys. Rev. A {\bf 69}, 042302 (2004).
\bibitem{Burkard2006}
G. Burkard and A. Imamoglu,
Phys. Rev. B {\bf 74}, 041307({\bf R}) (2006).
\bibitem{RWA}
Cohen-Tannoudji C., Dupont-Roc J., G.Grynberg, {\it Atom-Photon Interactions}, John Wiley and Sons, 
INC, 1992.
\bibitem{JaynesCummings}
E. T. Jaynes and F. W. Cummings,
Proc. IEEE {\bf 51}, 89 (1963).
\bibitem{Blais2007}
A. Blais, J. Gambetta, A. Wallraff, D. I. Schuster, S. M. Girvin, M. H. Devoret, and  R. J. 
Schoelkopf,
Phys. Rev. A {\bf 75}, 032329 (2007).
\bibitem{Vitaly}
V. N. Golovach, A. Khaetskii, and D. Loss,
Phys. Rev. Lett. {\bf 93}, 016601 (2004).

\bibitem{Rashba2001}
L. S. Levitov and E. I. Rashba,
Phys. Rev. B {\bf 67}, 115324 (2003).
\bibitem{Flindt2006}
C. Flindt, A. S. Sorensen and K. Flensberg,
Phys. Rev. Lett. {\bf 97} 240501 (2006).
\bibitem{Norihiko1}
N. Nishiguchi,
Phys. Rev. B, {\bf 50}, 10970 (1994).
\bibitem{book}
A. N. Cleland, {\it Foundations of Nanomechanics.  From Solid-State Theory to Devive Applications}, 
Springer-Verlag Berlin Heidelberg, 2003.
\bibitem{Norihiko2}
N. Nishiguchi, Phys. Rev. B. {\bf 52}, 5279 (1995).
\bibitem{normalization}
M. A. Stroscio, K. W. Kim, SeGi Yu, and A. Ballato,
J. Appl. Phys, {\bf 76} (8), 4670 (1994).
\bibitem{Massoud}
M. Borhani,  V. N. Golovach, and D. Loss,
Phys. Rev. B {\bf 73}, 155311 (2006).
\bibitem{Weiss}
U. Weiss, {\it Quantum Dissipative Systems}, World Scientific, Second Edition (1998).
\bibitem{Meunier}
T. Meunier, I. T. Vink, L. H. van Beveren, K-J. Tielrooij, R. Hanson, F. H. Koppens, H. P. Tranitz, W. Wegscheider, L. P. Kouwenhoven, and L. M. Vandersypen,
Phys. Rev. Lett. {\bf 98}, 126601 (2007).
\bibitem{Khaetskii2003}
A. Khaetskii, D. Loss, and L. Glazman,
Phys. Rev. Lett. {\bf 88}, 186802 (2002).
\bibitem{Coish2004}
W. A. Coish and D. Loss, Phys. Rev. B,{\bf 70},195340 (2004).
\bibitem{Klauser}
D. Klauser, W. A. Coish, and D. Loss,
Phys. Rev. B,{\bf 73}, 205302 (2006).
\bibitem{Stepanenko}
D. Stepanenko, G. Burkard, G. Giedke, and A. Imamoglu,
Phys. Rev. Lett. {\bf 96}, 136401 (2006).
\bibitem{Lukin2006}
T. Gaebel et al.,
Nature Phys. {\bf 2}, 408 (2006).
\bibitem{HansonNV2006}
R. Hanson, F. Mendoza, R. J. Epstein, and  D. D. Awschalom,
Phys. Rev. Lett. {\bf 97}, 087601 (2006).
\bibitem{Loss2001}
M. N. Leuenberger and D. Loss,
Nature (London) {\bf 410}, 789 (2001).
\bibitem{Tejada2001}
J. Tejada, E. M. Chudnovsky, E. del Barco, J. M. Hernandez, and T. P. Spiller,
Nanotechnology {\bf 12}, 2  181 (2001).
\bibitem{Lehmann2007}
J. Lehmann, A. Gaita-Ari\~{n}o, E. Coronado, and D. Loss,
Nature Nanotech. {\bf 2}, 312 (2007).
\bibitem{Filippo2006}
F. Troiani, A. Ghirri, M. Affronte, S. Carretta, P. Santini, G. Amoretti, S. Piligkos, G. Timco, 
and  R. E. P. Winpenny,
Phys. Rev. Lett. {\bf 94}, 207208 (2005).



\end{references}
\end{document}